\documentclass{IEEEtran}
\usepackage[utf8]{inputenc}
\usepackage{cite}
\usepackage{rotating}
\usepackage{booktabs}
\usepackage[usenames, dvipsnames]{color}
\usepackage[T1]{fontenc}
\usepackage{amsmath}
\usepackage[table]{xcolor}
\usepackage{epsfig,endnotes}
\usepackage[colorinlistoftodos,prependcaption,textsize=tiny]{todonotes}
\usepackage{hyperref}
\hypersetup{
  colorlinks,
  linkcolor={green!80!black},
  citecolor={red!70!black},
  urlcolor={blue!70!black}
}
\usepackage{breakurl}

\usepackage{mathtools}
\usepackage{url}
\usepackage[table]{xcolor}
\usepackage[noend]{algpseudocode}
\usepackage{tabularx,colortbl}
\usepackage{makecell}
\usepackage{multirow}
\usepackage[Symbolsmallscale]{upgreek}
\usepackage{algorithm}
\usepackage{algpseudocode}
\usepackage{mathtools}
\usepackage{url}
\usepackage{authblk}
\usepackage{tablefootnote}
\usepackage{siunitx}
\usepackage{listings}
\definecolor{keywords}{RGB}{255,0,90}
\definecolor{comments}{RGB}{0,0,113}
\usepackage{amssymb}
\usepackage{pifont}
\usepackage{bbding}
\usepackage{balance}
\usepackage{soul}
\usepackage{xurl}

\makeatletter
\newcommand\notsotiny{\@setfontsize\notsotiny{7}{8}}
\makeatother

\usepackage{subcaption}
\DeclareCaptionSubType * [alph]{table}
\def\tablename{Table}
\def\figurename{Figure}

\newcommand{\vthead}[1]{\makebox[1em][l]{\rotatebox{90}{#1}}}

\definecolor{lightgray}{gray}{0.95}
\definecolor{darkgray}{gray}{0.75}
\newcolumntype{?}{!{\color{darkgray}\vrule width 0.4pt}}

\newcommand{\pie}[1]{
  \hspace{-.8ex}\hskip-\tabcolsep
  \begin{tikzpicture}[scale=0.8, baseline=-.5ex]
    \draw (0,0) circle (1ex);
    \fill[white] (0,0) circle (1ex);
    \ifthenelse{\equal{#1}{0}}{
    }{
      \fill (0,1ex) arc (90:int(#1*90+90):1ex) -- (0,0) -- cycle;
    }
  \end{tikzpicture}
  \hspace{-1.6ex}\hskip-\tabcolsep
}
\newcommand{\piel}[1]{\hspace{3pt}\pie{#1}\hspace{5pt}}

\newcommand{\unsure}{{\textbf{?}}}

\newcommand{\cmark}{\ding{51}}

\usepackage{tcolorbox}
\newtcolorbox{mtbox}[1]{left=0.25mm, right=0.25mm, top=0.25mm, bottom=0.25mm, colframe=red!50!black, boxrule=0.5pt, title={#1}, fonttitle=\bfseries, coltitle=red!50!black, attach title to upper={\ --\ }}

\usepackage[framemethod=tikz]{mdframed}
\mdfdefinestyle{remarkstyle}{
backgroundcolor=lightgray,
roundcorner=3pt,
innerleftmargin=3pt,
innerrightmargin=3pt,
innertopmargin=3pt,
innerbottommargin=3pt,
linewidth=0.5pt,
}

\newcounter{gap}[section]
\newenvironment{gap}{\refstepcounter{gap}
\begin{mdframed}[style=remarkstyle]
\noindent \textbf{Scientific Gap~\thegap}: \em
}
{
\end{mdframed}
\vspace{-1mm}
}

\newcommand{\nsection}[1]{\vspace{-0.1cm}\section{#1}\vspace{-0.05cm}}
\newcommand{\nsubsection}[1]{\vspace{-0.15cm}\subsection{#1}\vspace{-0.05cm}}

\begin{document}

\title{\huge SoK: On the Semantic AI Security in Autonomous Driving}

\author{
Junjie Shen, Ningfei Wang, Ziwen Wan, Yunpeng Luo, Takami Sato, Zhisheng Hu\IEEEauthorrefmark{2}, Xinyang Zhang\IEEEauthorrefmark{3}, \mbox{Shengjian Guo\IEEEauthorrefmark{3}}, Zhenyu Zhong\IEEEauthorrefmark{3}, Kang Li\IEEEauthorrefmark{3}, Ziming Zhao\IEEEauthorrefmark{4}, Chunming Qiao\IEEEauthorrefmark{4}, Qi Alfred Chen \\
\{junjies1, ningfei.wang, ziwenw8, yunpel3, takamis, alfchen\}@uci.edu, \IEEEauthorrefmark{2}zhu@zoox.com \\
\IEEEauthorrefmark{3}\{xinyangzhang, sjguo, edwardzhong, kangli01\}@baidu.com, \IEEEauthorrefmark{4}\{zimingzh, qiao\}@buffalo.edu\\
UC Irvine, \IEEEauthorrefmark{2}ZOOX, \IEEEauthorrefmark{3}Baidu Security, \IEEEauthorrefmark{4}University at Buffalo}

\maketitle

\thispagestyle{plain}
\pagestyle{plain}

\begin{abstract}

Autonomous Driving (AD) systems rely on AI components to make safety and correct driving decisions. Unfortunately, today's AI algorithms are known to be generally vulnerable to adversarial attacks. However, for such AI component-level vulnerabilities to be semantically impactful at the system level, it needs to address non-trivial semantic gaps both (1) from the system-level attack input spaces to those at AI component level, and (2) from AI component-level attack impacts to those at the system level. In this paper, we define such research space as \textit{semantic AI security} as opposed to generic AI security. Over the past 5 years, increasingly more research works are performed to tackle such semantic AI security challenges in AD context, which has started to show an exponential growth trend. However, to the best of our knowledge, so far there is no comprehensive systematization of this emerging research space.

In this paper, we thus perform the first systematization of knowledge of such growing semantic AD AI security research space. In total, we collect and analyze 53 such papers, and systematically taxonomize them based on research aspects critical for the security field such as the attack/defense targeted AI component, attack/defense goal, attack vector, attack knowledge, defense deployability, defense robustness, and evaluation methodologies. We summarize 6 most substantial scientific gaps observed based on quantitative comparisons both vertically among existing AD AI security works and horizontally with security works from closely-related domains. With these, we are able to provide insights and potential future directions not only at the design level, but also at the research goal, methodology, and community levels. To address the most critical scientific methodology-level gap, we take the initiative to develop an open-source, uniform, and extensible system-driven evaluation platform, named \textit{PASS}, for the semantic AD AI security research community. We also use our implemented platform prototype to showcase the capabilities and benefits of such a platform using representative semantic AD AI attacks.

\end{abstract}
\vspace{-0.2cm}
\nsection{Introduction} \label{sec:intro}

Autonomous Driving (AD) vehicles are now a reality in our daily life, where a wide variety of commercial and private AD vehicles are already driving on the road. For example, commercial AD services, such as self-driving taxis~\cite{baidu_driverless_robotaxi, waymo-one}, buses~\cite{baidu_bus_service, uk_bus}, trucks~\cite{tusimple-truck, aurora_truck}, delivery vehicles~\cite{nuro_charge_delivery_service} are already publicly available, not to mention the millions of Tesla cars~\cite{tesla_sold_2_million_cars} that are equipped with Autopilot~\cite{news_tesla_autopilot}. To achieve driving autonomy in complex and dynamic driving environments, AD systems are designed with a collection of AI components to handle the core decision-making process such as perception, localization, prediction, and planning, which essentially forms the ``brain'' of the AD vehicle. However, this makes these AI components highly security-critical as errors in them can cause various road hazards and even fatal consequences~\cite{tesla2020wsj, uber_crash}.

Unfortunately, today's AI algorithms, especially deep learning, are known to be generally vulnerable to adversarial attacks~\cite{szegedy2014intriguing, goodfellow2014explaining}. However, since these AI algorithms are only components of the entire AD system, it is widely recognized that such generic AI component-level vulnerabilities do not necessarily lead to system-level vulnerabilities~\cite{seshia2020semantic, pierazzi2020intriguing, dreossi2019compositional, jia2020fooling}. This is mainly due to the large \textbf{semantic gaps}: (1) from the system-level attack input spaces (e.g., adding stickers~\cite{eykholt2018physical}, laser shooting~\cite{cao2019adversarial}) to those at the AI component level (e.g., image pixel changes~\cite{szegedy2014intriguing, goodfellow2014explaining}), or \textit{system-to-AI semantic gap}, which needs to overcome fundamental design challenges to map successful attacks at the AI input space back to the problem space, generally called the \textit{inverse-feature mapping problem}~\cite{pierazzi2020intriguing}; and (2) from AI component-level attack impacts (e.g., misdetected road objects) to those at the system level (e.g., vehicle collisions), or \textit{AI-to-system semantic gap}, which is also quite non-trivial, e.g., when the misdetected object is at a far distance for automatic emergency braking~\cite{dreossi2019compositional, seshia2020semantic}, or the misdetection can be tolerated by subsequent AI modules like object tracking~\cite{jia2020fooling}. Thus, for an AI security work to be semantically meaningful at the system level, it must explicitly or implicitly address these 2 general semantic gaps. In this paper, we call such research space \textbf{semantic AI security} (as opposed to generic AI security), following the semantic adversarial deep learning concept by Seshia et al.~\cite{seshia2020semantic}. 

Over the past 5 years, increasingly more research works are performed to tackle the aforementioned semantic AI security challenges in AD context, which started to show an exponential growth trend since 2019 (Fig.~\ref{fig:stats_paper_count}). However, to the best of our knowledge, so far there is no comprehensive systematization of this emerging research space. There are surveys related to AD security, but they either did not focus on AD AI components (e.g., on sensor/hardware~\cite{kim2021cybersecurity}, in-vehicle network~\cite{ren2019security}), or touched upon AD AI components but did not focus on the works that addressed the semantic AI security challenges above~\cite{deng2021deep, qayyum2020securing}. Since (1) the latter ones are much more semantically and thus practically meaningful for AD systems, (2) now a substantial amount of them have appeared (over 50 as in Fig.~\ref{fig:stats_paper_count}), and (3) such attacks in AD context have especially high safety-criticality since AD vehicles are heavy, fast-moving, and operate in public spaces, we believe now is a good time to summarize the current status, trends, as well as scientific gaps, insights, and future research directions.

\begin{table*}[tbp]
\footnotesize
\begin{minipage}{0.31\linewidth}
	\centering
	\vspace{-0.03cm}
    \includegraphics[width=.87\columnwidth]{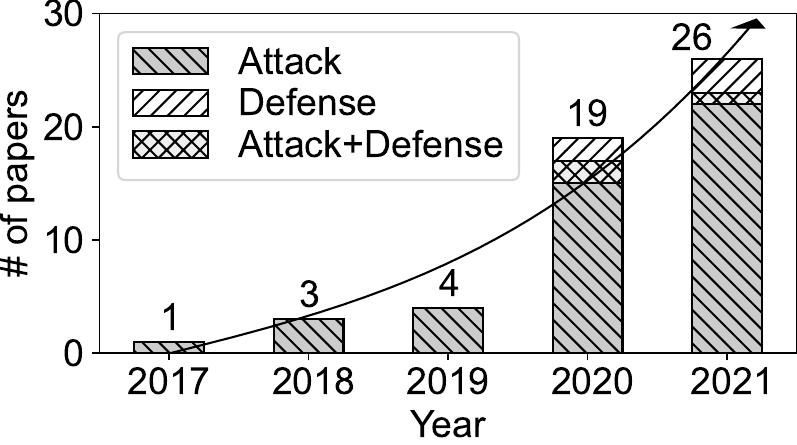}
    \vspace{-0.17cm}
    \captionof{figure}{Number and trends of semantic AD AI security papers (collected with a focus on top-tier venues (\S\ref{sec:background_scope})).}
    \label{fig:stats_paper_count}
\end{minipage}\hfill %
\begin{minipage}{0.67\linewidth}
	\centering
	\vspace{0.1cm}
    \includegraphics[width=\columnwidth]{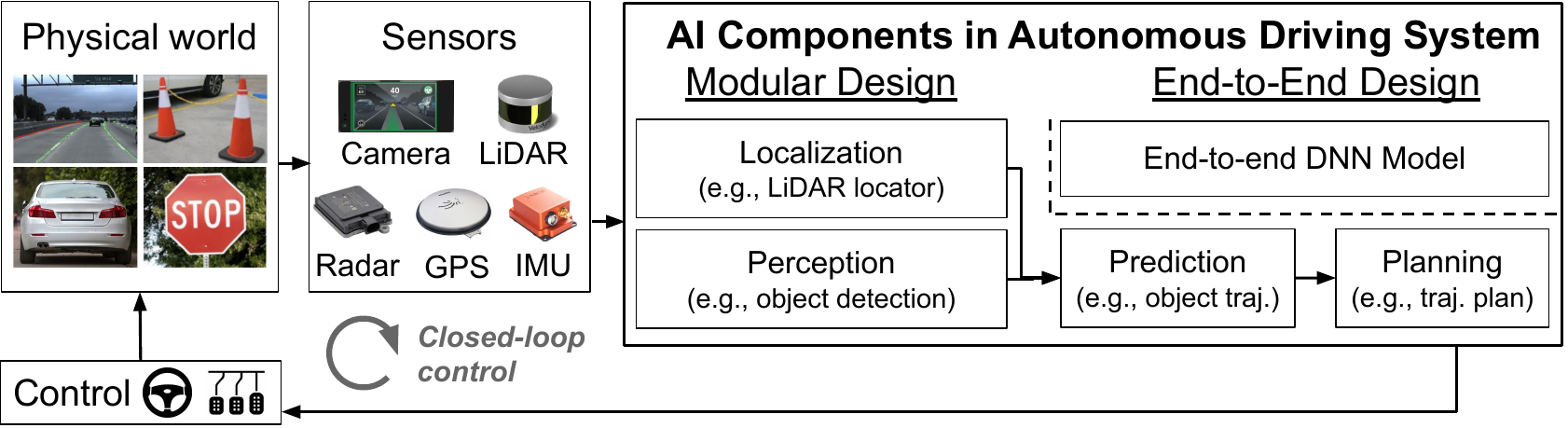}
    \captionof{figure}{Overview of AD system designs and the roles of AD AI components.}
    \label{fig:overview_ad}
\end{minipage}
\vspace{-0.2in}
\end{table*}

In this paper, we perform the first systematization of knowledge (SoK) of the growing semantic AD AI security research space. In total, we collect and analyze 53 such papers, with a focus on those published in (commonly-recognized) top-tier venues across security, computer vision, machine learning, AI, and robotics areas in the past 5 years since the first one appeared in 2017 (\S\ref{sec:background_scope}). Next, we taxonomize them based on research aspects critical for the security field, including the targeted AI component, attack/defense goal, attack vector, attack knowledge, defense deployability, defense robustness, as well as evaluation methodologies (\S\ref{sec:sok}). For each research aspect, we emphasize the observed domain/problem-specific design choices, and summarize their current status and trends. 

Based on the systematization, we summarize 6 most substantial scientific gaps (\S\ref{sec:gaps}) observed based on quantitative comparisons both vertically among existing AD AI security works and horizontally with security works from closely-related domains (e.g., drone~\cite{nassi2021sok}). With these, we are able to provide insights and potential future directions not only at the design level (e.g., under-explored attack goals and vectors), but also at the research goal and methodology levels (e.g., the general lack of system-level evaluations), as well as at the community level (e.g., the substantial lacking of open-sourcing specifically for the works in the security community).

Among all these scientific gaps, the one on the general lack of system-level evaluation is especially critical as it may lead to meaningless attack/defense progress at the system level due to the AI-to-system semantic gap (\S\ref{sec:gaps_system_level_eval}). To effectively fill this gap, it is highly desired to have a community-level effort to collectively build a common system-level evaluation infrastructure, since (1) the engineering efforts for building such infrastructure share common design/implementation patterns; and (2) in AD context, the system-level evaluation results are only comparable (and thus scientifically-meaningful) if the same evaluation scenario and metric calculation are used.

In this paper, we thus take the initiative to address this critical scientific methodology-level gap by developing a uniform and extensible system-driven evaluation platform, named \textit{PASS} (\underline{P}latform for \underline{A}utonomous driving \underline{S}afety and \underline{S}ecurity), for the semantic AD AI security research community (\S\ref{sec:platform}). We choose a simulation-centric hybrid design leveraging both simulation and real vehicles to balance the trade-offs among fidelity, affordability, safety, flexibility, efficiency, and reproducibility. The platform will be fully open-sourced (will be available on our project website~\cite{sok_website}) so that researchers can collectively develop new interfaces to fit future needs, and also contribute attack/defense implementations to form a semantic AD AI security benchmark, which can improve comparability, reproducibility, and also encourage open-sourcing. We have implemented a prototype, and use it to showcase an example usage that performs system-level evaluation of the most popular AD AI attack category, camera-based STOP sign detection~\cite{eykholt2018physical, zhao2019seeing}, using 45 different combinations of system-level scenario setups (i.e., speeds, weather, and lighting). We find that the AI component-level and AD system-level results are quite different and often contradict each other in common driving scenarios, which further demonstrates the necessity and benefits of such system-level evaluation infrastructure building effort. Demos are available on our project website \textbf{\url{https://sites.google.com/view/cav-sec/pass}}~\cite{sok_website}.

In summary, this work makes the following contributions:
\begin{itemize}
    \item We perform the first SoK of the growing semantic AD AI security research space. In total, we collect and analyze 53 such papers, and systematically taxonomize them based on research aspects critical for the security field, including the targeted AI component, attack/defense goal, attack vector, attack knowledge, defense deployability, defense robustness, and evaluation methodologies.
    \item We summarize 6 most substantial scientific gaps observed based on quantitative comparisons both vertically among existing AD AI security works and horizontally with security works from closely-related domains. With these, we are able to provide insights and potential future directions not only at the design level, but also at the research goal, methodology, and community levels.
    \item To address the most critical scientific methodology-level gap, we take the initiative to develop an open-source, uniform, and extensible system-driven evaluation platform \textit{PASS} for the semantic AD AI security research community. We also use our implemented prototype to showcase the capabilities and benefits of such a platform using representative AD AI attacks. 
\end{itemize}
\nsection{Background and Systematization Scope} \label{sec:background}
\vspace{0.1in}

\nsubsection{A Primer on AD Systems and the AI Components} \label{sec:background_ad_ai}

\subsubsection{AD Levels and Deployment Status} \label{sec:background_ad_levels}
The Society of Automotive Engineers (SAE) defines 6 AD levels -- Level 0 (L0) to 5 (L5)~\cite{sae2018}, where as the level increasing, the driving is more automated. In particular, L0 refers to no automation. However, the vehicle may provide warning features such as Forward Collision Warning and Lane Departure Warning. L1 refers to driver assistance and is the minimum level of AD. In L1 vehicles, the AD system is in charge of \textit{either} steering \textit{or} throttling/braking. Examples of L1 features are Automated Lane Centering and Adaptive Cruise Control. L2 means partial automation, where the AD system controls \textit{both} steering \textit{and} throttling/braking. Although L1 and L2 can at least partially drive the vehicle, the driver must actively monitor and be ready to take over at any circumstances. When the autonomy level goes beyond L3, the driver does not need to be attentive when the AD system is operating in its Operational Design Domains (ODDs). However, in L3, the driver is required to take over when the AD system requests to do so. None of L4 and L5 vehicles require a driver seat. The difference is that L4 AD systems can only operate in limited ODDs whereas L5 can handle all possible driving scenarios. Currently, L2 AD is already widely available and targets consumer vehicles (e.g., Tesla Autopilot~\cite{news_tesla_autopilot}, Cadillac Super Cruise~\cite{news_super_cruise}, OpenPilot~\cite{news_openpilot}). L4 AD is under rapid deployment targeting transportation services such as self-driving taxis~\cite{baidu_driverless_robotaxi, waymo-one}, buses~\cite{baidu_bus_service, uk_bus}, trucks~\cite{tusimple-truck, aurora_truck}, and delivery robots~\cite{nuro_charge_delivery_service}, with some of them already entered the commercial operation stage that charges customers~\cite{baidu_first_paid_taxi_service, waymo-taxi}.

\subsubsection{AI Components in AD Systems} \label{sec:background_ai_components} 
The AD systems generally have multiple AI components, including perception, localization, prediction, and planning as the major categories, as shown in Fig.~\ref{fig:overview_ad}. 
\textit{Perception} refers to perceiving the surrounding environments and extracting the semantic information for driving, such as road object (e.g. pedestrian, vehicles, obstacles) detection, object tracking, segmentation, lane/traffic light detection. Perception module usually takes diverse sensor data as the inputs, including camera, LiDAR, RADAR, etc. \textit{Localization} refers to finding the position of the AD vehicle relative to a reference frame in an environment~\cite{kuutti2018survey}. \textit{Prediction} aims to estimate the future status (position, heading, velocity, acceleration, etc.) of surrounding objects. \textit{Planning} aims to make trajectory-level driving decisions (e.g., cruising, stopping, lane changing, etc.) that are safe and correct (e.g., conforming to traffic rules). Besides such modular design, people also explored \textit{end-to-end DNN models} for AD. Currently, since the modular design is easier to debug, interpret, and hard-code safety rules/measures, it is predominately adopted in industry-grade AD systems, while the end-to-end designs are only for demonstration purposes~\cite{yurtsever2020survey}. For more details, we refer the readers to recent surveys~\cite{paden2016survey, yurtsever2020survey, kuutti2018survey, tampuu2020survey}.

\nsubsection{Adversarial (AI) Attacks and Defenses} \label{sec:background_aml}
Recent works find that AI models are generally vulnerable to \textit{adversarial attacks}~\cite{szegedy2014intriguing, carlini2017towards}
Some works further explored such attacks in the physical world~\cite{chen2018shapeshifter, eykholt2018physical, cao2019adversarial, cao2021invisible}. With that, some defenses~\cite{Xu:2018:ndss, lecuyer2019certified} are proposed recently to defend against such attacks. Such AI attacks/defenses are directly related to AD systems due to the high reliance on AI components (\S\ref{sec:background_ai_components}). However, as described in~\S\ref{sec:intro}, generic AI component-level vulnerabilities do not necessarily lead to system-level vulnerabilities due to the general system-to-AI and AI-to-system semantic gaps. In this paper, we thus focus on the works that address \textit{semantic AI security} in the AD context.

\nsubsection{Systematization Scope} \label{sec:background_scope}

In our systematization, we consider within-scope the attacks that aim to address the \textit{semantic AI security} challenges (defined in~\S\ref{sec:intro}) and the defenses that are designed specifically for addressing the AI component-level vulnerabilities revealed in such attack works. Thus, we consider out-of-scope the works that assume digital space perturbations without justifying the feasibility in the AD context (i.e., without addressing the system-to-AI semantic gap)~\cite{pei2017deepxplore, tian2018deeptest, guo2018dlfuzz, xie2017adversarial} and the ones that focus only on AI algorithms that are not used in representative AD system designs today (\S\ref{sec:background_ai_components}), e.g., general image classification~\cite{kurakin2016adversarial, dong2018boosting, chen2017zoo}. Curious readers can refer to general adversarial attack/defense SoK~\cite{papernot2018sok} and surveys~\cite{akhtar2018threat, chakraborty2018adversarial}.

When collecting the papers, we mainly focus on the ones published in commonly-recognized top-tier venues~\cite{csrankings} in closely-related fields to AD AI (i.e., security, Computer Vision (CV), Machine Learning (ML), AI, and robotics), as well as a few well-known works published in arXiv and other venues based on our best knowledge. Particularly, for the top-tier venues, we \textit{exhaustively} search over the paper lists from 2017 to 2021 to find the ones that fall into our scope above.

\nsubsection{Related Work} \label{sec:related_work}

Before this paper, a few AD security-related surveys have been published~\cite{deng2021deep, qayyum2020securing, kim2021cybersecurity, ren2019security}, but none of them focus on the emerging semantic AD AI research space (i.e., our scope defined in~\S\ref{sec:background_scope}). For example, Kim et al.~\cite{kim2021cybersecurity} and Ren et al.~\cite{ren2019security} focus on AD-related sensor/hardware security and in-vehicle network security, instead of AD AI components. Qayyum et al.~\cite{qayyum2020securing} and Deng et al.~\cite{deng2021deep} touched upon the security of AI components in AD, but did not focus on the works that addressed the semantic AI security challenges (e.g., most of the included works are on generic AI and sensor security without studying impacts on AD AI behaviors). In fact, Deng et al.~\cite{deng2021deep} considers the semantic AD AI security as a future direction. This is mainly because both of them focus on works published in 2019 and before. However, the majority (85\%) and the exponential growth of semantic AD AI security works are after 2019 (Fig.~\ref{fig:stats_paper_count}). This leads to much more complete, diverse, and quantifiable observations of the current status, trends, and scientific gaps of this emerging research space in this paper, not to mention that we also take the initiative to address one of the most critical gaps.

In the general CPS (Cyber-Physical System) area, there are also SoKs on the security of technologies related to AD AI, for example on drones~\cite{nassi2021sok} that share similar controller designs, on Automatic Speech Recognition and Speaker Identification (ASR/SI)~\cite{abdullah2021sok} that are also AI-enabled CPS, and on sensor technology~\cite{yan2020sok} that provide the main inputs to AD AI. In comparison, the SoKs on drone and ASR/SI security focused on domain-specific attack vectors (e.g., ground control station channel and voice signals), which thus lead to vastly different set of systematized knowledge (\S\ref{sec:sok}) due to the CPS domain differences. The SoK on sensor security is complementary to us as it does not directly consider AI component security but sensor attack is one major attack vector to AD AI (\S\ref{sec:sok_attack_vectors}).
\vspace{-0.1cm}
\nsection{Systematization of Knowledge} \label{sec:sok}

In this section, we taxonomize the semantic AD AI security works published in the past 5 years since the first one appeared in 2017. In total, 53 papers fall into our scope (\S\ref{sec:background_scope}) across security, CV, ML, AI, and robotics research areas; 48 discovered new attacks (Table~\ref{tab:attacks}) and 8 developed effective new defense solutions (Table~\ref{tab:defenses}). Fig.~\ref{fig:stats_paper_count} shows the number of papers in each year. The earliest paper~\cite{lu2017no} dates back to 2017 when adversarial attacks (\S\ref{sec:background_aml}) began to be applied to many real-world application scenarios including AD. Since then, AD AI security has gained much attention as reflected in the exponential growth trend of paper numbers over the years.

Similar to many other fields, AD AI security research also has gone through enlightening periods where early beliefs are later overturned. For example, the 2017 paper~\cite{lu2017no} questions the severity of adversarial attacks in AD vehicles and claims that ``no need to worry about adversarial examples in object detection in autonomous vehicles''. Soon after that, two papers in 2018 propose successful attacks against camera object detection with physical-world attack demonstrations. As the community exponentially grew after 2019 and now at over 50 papers in this research space, we think now might be a good time to systematize the existing research efforts.

\vspace{-0.1cm}
\nsubsection{(Attack/Defense) Targeted AI Components} \label{sec:sok_target_ai_components}

\textit{Status and trends.}  The targeted AI components in the existing works are summarized in Tables~\ref{tab:attacks} and~\ref{tab:defenses}. As shown in the tables and Fig.~\ref{fig:stats_targeted_ai_components} in Appendix, most ($>$86\%) of the existing works target perception, while localization, chassis, and end-to-end driving are all less or equal to 6.2\%. Among the perception works, the two most popular ones are camera (60.0\%) and LiDAR (21.5\%) perception. More detailed summary of their designs, applications in AD systems, and vulnerabilities are in Appendix~\ref{appendix:sok_target_ai_components}. Currently, \textit{none} of the existing works study the downstream AI components such as prediction and planning, which will be discussed more in \S\ref{sec:gaps_downstream_components}.

\vspace{-0.1cm}
\nsubsection{Systematization of Semantic AD AI Attacks} \label{sec:sok_attacks}

Table~\ref{tab:attacks} summarizes the semantic AD AI attacks. We taxonomize them based on 3 research aspects critical for the security field: Attack goal, attack vector, and attacker's knowledge.

\begin{table*}[]
\centering
\setlength{\tabcolsep}{3.5pt}
\begin{tabular}{|c?c|r|cc|ccc|ccc?cccccc?cc|c|cc|c|}
\cline{1-23}
  \multicolumn{2}{|c|}{\multirow{4}{*}{}} &
  \multicolumn{1}{c|}{\multirow{4}{*}{}} &
  \multicolumn{1}{c}{\multirow{4}{*}{}} &
  \multicolumn{1}{c|}{\multirow{4}{*}{}} &
  \multicolumn{3}{c|}{\multirow{3}{*}{\begin{tabular}[c]{@{}c@{}}Attack\\ goal\end{tabular}}} &
  \multicolumn{11}{c|}{Attack vector} &
  \multirow{4}{*}{} &
  \multicolumn{2}{c|}{\multirow{3}{*}{\begin{tabular}[c]{@{}c@{}}Eval.\\ level\end{tabular}}} &
  \multirow{4}{*}{} \\ \cline{9-19}
  \multicolumn{2}{|c|}{} &
  \multicolumn{1}{c|}{} &
  \multicolumn{1}{c}{} &
  \multicolumn{1}{c|}{} &
  \multicolumn{3}{c|}{} &
  \multicolumn{9}{c?}{Physical-layer} &
  \multicolumn{2}{c|}{\multirow{2}{*}{\begin{tabular}[c]{@{}c@{}}Cyber\\layer\end{tabular}}} &
  &
  \multicolumn{2}{c|}{} &
  \\ \cline{9-17}
  \multicolumn{2}{|c|}{} &
  \multicolumn{1}{c|}{} &
  \multicolumn{1}{c}{} &
  \multicolumn{1}{c|}{} &
  \multicolumn{3}{c|}{} &
  \multicolumn{3}{c?}{\begin{tabular}[c]{@{}c@{}}Phys. world\end{tabular}} &
  \multicolumn{6}{c?}{Sensor attack} &
  \multicolumn{2}{c|}{} &
  &
  \multicolumn{2}{c|}{} &
  \\ \cline{9-19}
  \multicolumn{2}{|c|}{\begin{tabular}[c]{@{}c@{}}Targeted AI component \end{tabular}} &
  \multicolumn{1}{c|}{Paper} &
  \multicolumn{1}{c}{\vthead{Year}} &
  \multicolumn{1}{c|}{\vthead{Field}} &
  \vthead{Integrity} &
  \vthead{Confidentiality} &
  \vthead{Availability} &
  \vthead{Object texture} &
  \vthead{Object shape} &
  \vthead{Object position} &
  \vthead{GPS spoofing} &
  \vthead{LiDAR spoofing} &
  \vthead{Radar spoofing} &
  \vthead{Laser/IR light} &
  \vthead{Acoustic signal} &
  \vthead{Translucent patch} &
  \vthead{ML backdoor} &
  \vthead{Malware \& s/w compromise} &
  \vthead{Attacker's knowledge} &
  \vthead{Component-level} &
  \vthead{System-level} &
  \vthead{Open source} \\ \hline
  &
  &
  Lu et al.~\cite{lu2017no} &
  '17 &
  V &
  \cmark &
  &
  &
  \cmark &
  &
  &
  &
  &
  &
  &
  &
  &
  &
  &
  \pie{0} &
  \cmark &
  & 
  \\
  &
  &
  Eykholt et al.~\cite{eykholt2018physical} &
  '18 &
  S &
  \cmark &
  &
  &
  \cmark &
  &
  &
  &
  &
  &
  &
  &
  &
  &
  &
  \pie{0} &
  \cmark &
  &
  \\
  &
  &
  Chen et al.~\cite{chen2018shapeshifter} &
  '18 &
  M &
  \cmark &
  &
  &
  \cmark &
  &
  &
  &
  &
  &
  &
  &
  &
  &
  &
  \pie{0} &
  \cmark &
  &
  \cmark
  \\
  & 
  & 
  Zhao et al.~\cite{zhao2019seeing} &  
  '19 &
  S &
  \cmark &
  &
  &
  \cmark &
  &
  &
  &
  &
  &
  &
  &
  &
  &
  &
  \pie{0} &  
  \cmark & 
  & 
  \\ 
  & 
  & 
  Xiao et al.~\cite{xiao2019meshadv} &  
  '19 &
  V &
  \cmark &
  &
  &
  \cmark &
  \cmark &
  &
  &
  &
  &
  &
  &
  &
  &
  &
  \pie{0} &  
  \cmark &
  &
  \\ 
  & 
  & 
  Zhang et al.~\cite{zhang2019camou} &  
  '19 &
  M &
  \cmark &
  &
  &
  \cmark &
  &
  &
  &
  &
  &
  &
  &
  &
  &
  &
  \pie{2} &  
  \cmark &
  &
  \cmark
  \\ 
  & 
  & 
  Nassi et al.~\cite{nassi2020phantom} &  
  '20 &
  S &
  \cmark &
  &
  &
  \cmark &
  &
  &
  &
  &
  &
  &
  &
  &
  &
  &
  \pie{0} &  
  \cmark &
  \cmark &
  \\
  & 
  & 
  Man et al.~\cite{man2020ghostimage} &  
  '20 &
  S &
  \cmark &
  &
  &
  &
  &
  &
  &
  &
  &
  \cmark &
  &
  &
  &
  &
  \pie{0} &
  \cmark &
  &
  \cmark
  \\ 
  & 
  & 
  Hong et al.~\cite{hong2020avguardian} &
  '20 &
  S &
  \cmark & 
  & 
  & 
  & 
  & 
  & 
  & 
  & 
  & 
  & 
  & 
  & 
  & 
  \cmark & 
  \pie{0} & 
  & 
  \cmark & 
  \\ 
  & 
  & 
  Huang et al.~\cite{huang2020universal} & 
  '20 &  
  V &  
  \cmark &  
  &  
  &  
  \cmark &  
  &  
  &  
  &  
  &  
  &  
  &  
  &  
  &  
  &  
  &  
  \pie{0} &  
  \cmark &  
  &  
  \cmark 
  \\
  & 
  & 
  Wu et al.~\cite{wu2020making} &  
  '20 &  
  V &  
  \cmark &  
  &  
  &  
  \cmark &  
  &  
  &  
  &  
  &  
  &  
  &  
  &  
  &  
  &  
  &  
  \pie{0} &  
  \cmark &  
  &  
  \\ 
  & 
  Object & 
  Xu et al.~\cite{xu2020adversarial} &  
  '20 &  
  V &  
  \cmark &  
  &  
  &  
  \cmark &  
  &  
  &  
  &  
  &  
  &  
  &  
  &  
  &  
  &  
  &  
  \pie{0} &  
  \cmark &  
  &  
  \\
  & 
  detection & 
  Hu et al.~\cite{hu2020cca} &  
  '20 &  
  V &  
  \cmark &  
  &  
  &  
  \cmark &  
  &  
  &  
  &  
  &  
  &  
  &  
  &  
  &  
  &  
  &  
  \pie{2} &  
  \cmark &  
  &  
  \\
  & 
  & 
  Hamdi et al.~\cite{hamdi2020sada} &  
  '20 &  
  M &  
  \cmark &  
  &  
  &  
  &  
  \cmark &  
  &  
  &  
  &  
  &  
  &  
  &  
  &  
  &  
  &  
  \pie{2} &  
  \cmark &  
  &  
  \\ 
  & 
  & 
  Ji et al.~\cite{ji2021poltergeist} &  
  '21 &  
  S &  
  \cmark &  
  &  
  &  
  &  
  &  
  &  
  &  
  &  
  &  
  &  
  \cmark &  
  &  
  &  
  &  
  \pie{2} &  
  \cmark &  
  &  
  \\
  & 
  & 
  Lovisotto et al.~\cite{lovisotto2021slap} &  
  '21 &  
  S &  
  \cmark &  
  &  
  &  
  \cmark&  
  &  
  &  
  &  
  &  
  &  
  &  
  &  
  &  
  &  
  &  
  \pie{4} &  
  \cmark &  
  &  
  \cmark 
  \\ 
  & 
  & 
  Wang et al.~\cite{wang2021i} &  
  '21 &  
  S &  
  \cmark &  
  &  
  &  
  &  
  &  
  &  
  &  
  &  
  &  
  \cmark &  
  &  
  &  
  &  
  &  
  \pie{2} &  
  \cmark &  
  &  
  \\ 
  Camera & 
  & 
  K{\"o}hler et al.~\cite{kohler2021they} &  
  '21 &  
  S &  
  \cmark &  
  &  
  &  
  &  
  &  
  &  
  &  
  &  
  &  
  \cmark &  
  &  
  &  
  &  
  &  
  \pie{4} &  
  \cmark &  
  &  
  \\ 
  perception & 
  & 
  Wang et al.~\cite{wang2021daedalus} &  
  '21 &  
  S &  
  \cmark &  
  &  
  &  
  \cmark &  
  &  
  &  
  &  
  &  
  &  
  &  
  &  
  &  
  &  
  &  
  \pie{4} &  
  \cmark &  
  &  
  \cmark 
  \\ 
  & 
  & 
  Zolfi et al.~\cite{zolfi2021translucent} &  
  '21 &  
  V &  
  \cmark &  
  &  
  &  
  &  
  &  
  &  
  &  
  &  
  &  
  &  
  &  
  \cmark &  
  &  
  &  
  \pie{0} &  
  \cmark &  
  &  
  \\ 
  & 
  & 
  Wang et al.~\cite{wang2021dual} &  
  '21 &  
  V &  
  \cmark &  
  &  
  &  
  \cmark &  
  &  
  &  
  &  
  &  
  &  
  &  
  &  
  &  
  &  
  &  
  \pie{2} &  
  \cmark &  
  &  
  \cmark 
  \\ 
  & 
  & 
  Zhu et al.~\cite{zhu2021fooling} &  
  '21 &  
  M &  
  \cmark &  
  &  
  &  
  &  
  &  
  &  
  &  
  &  
  &  
  \cmark &  
  &  
  &  
  &  
  &  
  \pie{0} &  
  \cmark &  
  &  
  \\  \cline{2-23}  
  & 
  Semantic & 
  Nakka et al.~\cite{nakka2020indirect} &  
  '20 &  
  V &  
  \cmark &  
  &  
  &  
  \cmark &  
  &  
  &  
  &  
  &  
  &  
  &  
  &  
  &  
  &  
  &  
  \pie{0} &  
  \cmark &  
  &  
  \\ 
  & 
  segmentation & 
  Nesti et al.~\cite{nesti2022evaluating} & 
  '22 &  
  V &  
  \cmark &  
  &  
  &  
  \cmark &  
  &  
  &  
  &  
  &  
  &  
  &  
  &  
  &  
  &  
  &  
  \pie{0} &  
  \cmark &  
  &  
  \\  \cline{2-23}  
  & 
  & 
  Jha et al.~\cite{jha2020ml} &  
  '20 &  
  S &  
  \cmark &  
  &  
  &  
  &  
  &  
  &  
  &  
  &  
  &  
  &  
  &  
  &  
  &  
  \cmark &  
  \pie{0} &  
  &  
  \cmark &  
  \\ 
  & 
  Object & 
  Jia et al.~\cite{jia2020fooling} & 
  '20 &  
  M &  
  \cmark &  
  &  
  &  
  \cmark &  
  &  
  &  
  &  
  &  
  &  
  &  
  &  
  &  
  &  
  &  
  \pie{0} &  
  \cmark &  
  &  
  \cmark 
  \\ 
  & 
  tracking & 
  Ding et al.~\cite{ding2021towards} & 
  '21 &  
  M &  
  \cmark &  
  &  
  &  
  \cmark &  
  &  
  &  
  &  
  &  
  &  
  &  
  &  
  &  
  &  
  &  
  \pie{0} &  
  \cmark &  
  &  
  \\ 
  & 
  & 
  Chen et al.~\cite{chen2021unified} & 
  '21 &  
  M &  
  \cmark &  
  &  
  &  
  \cmark &  
  &  
  &  
  &  
  &  
  &  
  &  
  &  
  &  
  &  
  &  
  \pie{0} &  
  \cmark &  
  &  
  \\  \cline{2-23}  
  & 
  Lane &  
  Sato et al.~\cite{sato2021dirty} & 
  '21 &  
  S &  
  \cmark &  
  &  
  &  
  \cmark &  
  &  
  &  
  &  
  &  
  &  
  &  
  &  
  &  
  &  
  &  
  \pie{0} &  
  \cmark &  
  \cmark &  
  \cmark 
  \\ 
  &  
  detection &  
  Jing et al.~\cite{jing2021too} & 
  '21 &  
  S &  
  \cmark &  
  &  
  &  
  \cmark &  
  &  
  &  
  &  
  &  
  &  
  &  
  &  
  &  
  &  
  &  
  \pie{2} &  
  \cmark &  
  \cmark &  
  \\ \cline{2-23} 
  &  
  Traffic light&  
  Wang et al.~\cite{wang2021i} & 
  '21 &  
  S &  
  \cmark &  
  &  
  &  
  &  
  &  
  &  
  &  
  &  
  &  
  \cmark &  
  &  
  &  
  &  
  &  
  \pie{0} &  
  \cmark &  
  &  
  \\ 
  &  
  detection &  
  Tang et al.~\cite{tang2021fooling} & 
  '21 &  
  S &  
  \cmark &  
  &  
  &  
  &  
  &  
  &  
  \cmark &  
  &  
  &  
  &  
  &  
  &  
  &  
  &  
  \pie{0} &  
  &  
  \cmark &  
  \\ \hline
  &  
  &  
  Cao et al.~\cite{cao2019adversarial} &  
  '19 &  
  S &  
  \cmark &  
  &  
  &  
  &  
  &  
  &  
  &  
  \cmark &  
  &  
  &  
  &  
  &  
  &  
  &  
  \pie{0} &  
  \cmark &  
  \cmark &  
  \\ 
  &  
  &  
  Sun et al.~\cite{sun2020towards} & 
  '20 &  
  S &  
  \cmark &  
  &  
  &  
  &  
  &  
  &  
  &  
  \cmark &  
  &  
  &  
  &  
  &  
  &  
  &  
  \pie{4} &  
  \cmark &  
  &  
  \\ 
  & 
  & 
  Hong et al.~\cite{hong2020avguardian} & 
  '20 &
  S &
  \cmark & 
  & 
  & 
  & 
  & 
  & 
  & 
  & 
  & 
  & 
  & 
  & 
  & 
  \cmark & 
  \pie{0} & 
  & 
  \cmark & 
  \\ 
  &  
   &  
  Tu et al.~\cite{tu2020physically} & 
  '20 &  
  V &  
  \cmark &  
  &  
  &  
  &  
  \cmark &  
  &  
  &  
  &  
  &  
  &  
  &  
  &  
  &  
  &  
  \pie{0} &  
  \cmark &  
  &  
  \\ 
   &  
   Object&  
  Zhu et al.~\cite{zhu2021can} & 
  '21 &  
  S &  
  \cmark &  
  &  
  &  
  &  
  &  
  \cmark &  
  &  
  &  
  &  
  &  
  &  
  &  
  &  
  &  
  \pie{2} &  
  \cmark &  
  &  
  \\ 
  LiDAR &  
 detection &  
  Yang et al.~\cite{yang2021robust} & 
  '21 &  
  S &  
  \cmark &  
  &  
  &  
  &  
  \cmark &  
  &  
  &  
  &  
  &  
  &  
  &  
  &  
  &  
  &  
  \pie{2} &  
  \cmark &  
  \cmark &  
  \\ 
 perception &  
  &  
  Hau et al.~\cite{hau2021object} & 
  '21 &  
  S &  
  \cmark &  
  &  
  &  
  &  
  &  
  &  
  &  
  \cmark &  
  &  
  &  
  &  
  &  
  &  
  &  
  \pie{0} &  
  \cmark &  
  &  
  \\ 
  &  
  &  
  Li et al.~\cite{li2021fooling} & 
  '21 &  
  V &  
  \cmark &  
  &  
  &  
  &  
  &  
  &  
  \cmark &  
  &  
  &  
  &  
  &  
  &  
  &  
  &  
  \pie{0} &  
  \cmark &  
  &  
  \cmark 
  \\
  &  
  &  
  Zhu et al.~\cite{zhu2021adversarial} & 
  '21 &  
  O &  
  \cmark &  
  &  
  &  
  &  
  &  
  \cmark&  
   &  
  &  
  &  
  &  
  &  
  &  
  &  
  &  
  \pie{2} &  
  \cmark &  
  &  
  \\ \cline{2-23} 
  & 
  Semantic   &  
  Tsai et al.~\cite{tsai2020robust} & 
  '20 &  
  M &  
  \cmark &  
  &  
  &  
  &  
  \cmark &  
  &  
  &  
  &  
  &  
  &  
  &  
  &  
  &  
  &  
  \pie{0} &  
  \cmark &  
  &  
  \cmark 
  \\
  &  
  segmentation&  
  Zhu et al.~\cite{zhu2021adversarial} & 
  '21 &  
  O &  
  \cmark &  
  &  
  &  
  &  
  &  
  \cmark&  
   &  
  &  
  &  
  &  
  &  
  &  
  &  
  &  
  \pie{2} &  
  \cmark &  
  &  
  \\ \hline
  \multicolumn{1}{|c?}{RADAR perception} & 
  \multicolumn{1}{?c|}{Obj. detection} & 
  Sun et al.~\cite{sun2021control} & 
  '21 &  
  S &  
  \cmark &  
  &  
  &  
  &  
  &  
  &  
  &  
  &  
  \cmark &  
  &  
  &  
  &  
  &  
  &  
  \pie{2} &  
  \cmark &  
  \cmark &  
  \\ \hline
  \multicolumn{2}{|c|}{\multirow{2}{*}{\begin{tabular}[c]{@{}c@{}}MSF perception\end{tabular}}}
  &  
  Cao et al.~\cite{cao2021invisible} & 
  '21 &  
  S &  
  \cmark &  
  &  
  &  
  &  
  \cmark &  
  &  
  &  
  &  
  &  
  &  
  &  
  &  
  &  
  &  
  \pie{0} &  
  \cmark &  
  \cmark &  
  \cmark 
  \\ 
  \multicolumn{2}{|c|}{} 
  &  
  Tu et al.~\cite{tu2021exploring} & 
  '21 &  
  O &  
  \cmark &  
  &  
  &  
  &  
  \cmark &  
  &  
  &  
  &  
  &  
  &  
  &  
  &  
  &  
  &  
  \pie{0} &  
  \cmark &  
  &  
  \\ \hline
  \multicolumn{2}{|c|}{LiDAR localization } &  
  Luo et al.~\cite{luo2020stealthy} & 
  '20 &  
  S &  
  &  
  \cmark &  
  &  
  &  
  &  
  &  
  &  
  &  
  &  
  &  
  &  
  &  
  &  
  \cmark &  
  \pie{0} &  
  &  
  \cmark &  
  \\ 
  \multicolumn{2}{|c|}{MSF localization } &  
  Shen et al.~\cite{shen2020drift} & 
  '20 &  
  S &  
  \cmark &  
  &  
  &  
  &  
  &  
  &  
  \cmark &  
  &  
  &  
  &  
  &  
  &  
  &  
  &  
  \pie{0} &  
  \cmark &  
  \cmark &  
  \\
  \multicolumn{2}{|c|}{Camera localization } &  
  Wang et al.~\cite{wang2021i} & 
  '21 &  
  S &  
  \cmark &  
  &  
  &  
  &  
  &  
  &  
  &  
  &  
  &  
  \cmark &  
  &  
  &  
  &  
  &  
  \pie{0} &  
  \cmark &  
  &  
  \\ \hline
  \multicolumn{2}{|c|}{Chassis} & 
  Hong et al.~\cite{hong2020avguardian} &
  '20 &
  S &
  & 
  \cmark & 
  & 
  & 
  & 
  & 
  & 
  & 
  & 
  & 
  & 
  & 
  & 
  \cmark & 
  \pie{0} & 
  & 
  \cmark & 
  \\ \hline
  \multicolumn{2}{|c|}{\multirow{4}{*}{\begin{tabular}[c]{@{}c@{}}End-to-end driving\end{tabular}}} &
  Liu et al.~\cite{liu2018trojaning} &  
  '18 &  
  S &  
  \cmark &  
  &  
  &  
  \cmark &  
  &  
  &  
  &  
  &  
  &  
  &  
  &  
  &  
  \cmark &  
  &  
  \pie{0} &  
  \cmark &  
  \cmark &  
  \cmark 
  \\ 
  \multicolumn{2}{|c|}{} &  
  Kong et al.~\cite{kong2020physgan} & 
  '20 &  
  V &  
  \cmark &  
  &  
  &  
  \cmark &  
  &  
  &  
  &  
  &  
  &  
  &  
  &  
  &  
  &  
  &  
  \pie{0} &  
  \cmark &  
  &  
  \cmark 
  \\ 
  \multicolumn{2}{|c|}{} &  
  Hamdi et al.~\cite{hamdi2020sada} &  
  '20 &  
  M &  
  \cmark &  
  &  
  &  
  &  
  \cmark &  
  &  
  &  
  &  
  &  
  &  
  &  
  &  
  &  
  &  
  \pie{2} &  
  \cmark &  
  &  
  \\
  \multicolumn{2}{|c|}{} &  
  Boloor et al.~\cite{boloor2020attacking} &  
  '20 &  
  O &  
  \cmark &  
  &  
  &  
  \cmark  &  
  &  
  &  
  &  
  &  
  &  
  &  
  &  
  &  
  &  
  &  
  \pie{4} &  
  \cmark &  
  \cmark &  
  \cmark
  \\ \hline
\end{tabular}
\vspace{0.05in}

\footnotesize
Field: S = Security, V = Computer Vision, M = ML/AI, O = Others, e.g., Robotics, arXiv;\\
Attacker's knowledge:\  \piel{0} \ = white-box, \piel{2} \ = gray-box, \piel{4} \ = black-box

\caption[overview]{Overview of existing semantic AD AI attacks in our SoK scope (\S\ref{sec:background_scope}). (s/w = software)}
\label{tab:attacks}
\end{table*}

\subsubsection{Attack Goals} \label{sec:sok_attack_goals} 
We categorize the attack goals in the existing works based on the general security properties such as \textit{integrity}, \textit{confidentiality}, and \textit{availability}~\cite{stallings2012computer}:

\textbf{Integrity (of AI components).} 
Integrity in AD context can be viewed as the integrity of the AI component outputs (i.e., whether they are changed by the attacker), which can directly impact the correctness of the AI driving behaviors. From this view, its violations manifest as the following AD-specific attack goals considered in existing works:
\begin{itemize}
\vspace{-\topsep}
    \item \textit{Safety hazards.} Safety is the top priority in AD design~\cite{ad_safety_first}, and it is not only for the AD vehicle and its passengers, but also for other road users (e.g., other vehicles, pedestrians). Many existing attacks aim for safety hazards. For example, attacks on object detection and segmentation can cause safety hazards if a front vehicle is undetected~\cite{zhang2019camou, nakka2020indirect, wang2021dual}; attacks on object tracking can potentially lead to collisions if the front vehicle trajectory is incorrectly tracked~\cite{jia2020fooling, jha2020ml}; lane detection, localization, and end-to-end driving model attacks~\cite{sato2021dirty, jing2021too, shen2020drift, liu2018trojaning, kong2020physgan} can cause lane departure and thus potential collisions.
    \item \textit{Traffic rule violations.} Since AD systems by design are required to follow traffic rules, violations can lead to financial penalties for individuals and reputational losses for AD companies. Existing attacks aim to hide the STOP sign to cause STOP sign violations~\cite{chen2018shapeshifter, eykholt2018physical, zhao2019seeing}. Traffic light detection attack can cause the AD system to recognize a red light as green light, which may lead to red light violations~\cite{tang2021fooling}. In addition, the attacks on lane detection, localization, and end-to-end driving models can also cause vehicle lateral deviations, which can violate the lane line boundaries~\cite{sato2021dirty, jing2021too, shen2020drift, liu2018trojaning, kong2020physgan}.
    \item \textit{Mobility degradation.} A key benefit that AD technologies can bring is improved access to mobility-as-a-service (MaaS)~\cite{bertoncello2015ten}. A few works aim to degrade the mobility of AD vehicles by manipulating the AI component outputs. In those works, they fool either the object detection to recognize a static blocking obstacle~\cite{cao2019adversarial, sun2020towards, yang2021robust} or the traffic light detection to recognize a permanent red light~\cite{tang2021fooling}, which can cause the AD vehicle to be stuck on the road for a prolonged time. This not only delays the AD vehicle's trip to destination, but also may block the traffic and cause congestion.
\vspace{-\topsep}
\end{itemize}

\textbf{Confidentiality (Privacy).} Confidentiality is related to the sensitive information from or collected by the AD vehicle. This includes not only the vehicle identification information (e.g., the VIN from Chassis~\cite{hong2020avguardian}), but also the privacy-sensitive location data~\cite{luo2020stealthy} as it can reveal the passenger privacy.

\textbf{Availability (of AI components).} Availability in the general cybersecurity area means the systems, services, and information under viewed are accessible to users when they need them~\cite{stallings2012computer}. For an AD AI component, its ``users'' can be considered as all the downstream AD components and vehicle subsystems that are counting on its timely and reliable outputs to function correctly. Thus, the availability of an AD AI component can be defined as its capability to provide timely and reliable outputs. 

Following this definition, example attacks on AD AI availability can be attacks causing delays or failures in the outputting function of a given AI component, e.g., by interrupting its input/output messaging channels, causing software crashes/hangs in it, or by causing the vehicle system to fall back to human operators (so stop the outputting function of the whole AD AI stack).

\textit{Status and trends.} Vast majority (96.3\%) of existing works focus on integrity, while only 3.7\% are on confidentiality and so far none of them are on availability. More discussion on this are in \S\ref{sec:gaps_attack_goals}.

\subsubsection{Attack Vectors} \label{sec:sok_attack_vectors} 
Existing attacks on AD systems leverage a diverse set of attack vectors and we broadly categorize them in two categories: \textit{physical-layer} and \textit{cyber-layer}. Physical-layer attack vectors refer to those tampering the sensor inputs to the AI components via physical means. We further decompose physical-layer attack vectors  into \textit{physical-world attack} and \textit{sensor attack} vectors, where the former modifies the physical-world driving environment and the latter leverages unique sensor properties to inject erroneous measurements. Cyber-layer attack vector refers to those that require internal access to the AD system, its computation platform, or even the system development environment.

\textbf{Physical-world attack vectors:}
\begin{itemize}
\vspace{-\topsep}
    \item \textit{Object texture} refers to changing the surface texture (e.g., color) of 2D or 3D objects. It is often used by adversarial attacks to embed the malicious sensing inputs. In attack deployment, this is often fabricated as patches~\cite{sato2021dirty, eykholt2018physical, zhao2019seeing}, posters~\cite{lu2017no, chen2018shapeshifter, eykholt2018physical}, and camouflages~\cite{zhang2019camou, huang2020universal, hu2020cca}, or displayed by projectors~\cite{lovisotto2021slap} and digital billboards~\cite{nassi2020phantom}. Existing attacks have applied this attack vector on various objects such as STOP signs~\cite{lu2017no, eykholt2018physical, chen2018shapeshifter, zhao2019seeing}, road surfaces~\cite{sato2021dirty, jing2021too}, vehicles~\cite{zhang2019camou, huang2020universal, hu2020cca, wang2021dual}, clothes~\cite{wu2020making, xu2020adversarial}, physical billboards~\cite{kong2020physgan}. 
    To disguise as benign looking, a few attacks also constrain the texture perturbations to improve the attack stealthiness~\cite{huang2020universal, sato2021dirty, jing2021too}.
    \item \textit{Object shape} refers to perturbing the shape of 3D objects such as vehicles~\cite{xiao2019meshadv, tsai2020robust}, traffic cones~\cite{cao2021invisible}, rocks~\cite{cao2021invisible} or irregularly-shaped road objects~\cite{yang2021robust, tu2020physically, tu2021exploring}. Some were demonstrated in physical world via 3D printing~\cite{yang2021robust, cao2019adversarial}.
    \item \textit{Object position} refers to placing physical objects at specific locations in the environment. Prior work~\cite{zhu2021can} applies this attack vector by controlling a drone to hold a board at a particular location in the air to fool the LiDAR object detector to misdetect the front vehicle.
 \vspace{-\topsep}
\end{itemize}

\textbf{Sensor attack vectors:}

\begin{itemize}
\vspace{-\topsep}
    \item \textit{LiDAR spoofing} refers to injecting additional points in the LiDAR point cloud via laser shooting. Prior works carefully craft the injected point locations to fool the LiDAR object detection~\cite{cao2019adversarial, sun2020towards, hau2021object}.
    \item \textit{RADAR spoofing} refers to injecting malicious signals to RADAR inputs to cause it to resolve fake objects at specific distances and angles. Prior work~\cite{sun2021control} demonstrates RADAR spoofing capability in physical world and shows that it can cause AD system to recognize fake obstacles.
    \item \textit{GPS spoofing} refers to sending fake satellite signals to the GPS receiver, causing it to resolve positions that are specified by the attacker. Prior works leverage GPS spoofing to attack MSF localization~\cite{shen2020drift}, LiDAR object detection~\cite{li2021fooling}, and traffic light detection~\cite{tang2021fooling}.
    \item \textit{Laser/IR light} refers to injecting/projecting laser or light directly to the sensor rather than the environment. Prior works use such attack vector to project malicious light spots in the camera image such that it can misguide the camera localization~\cite{wang2021i} or object detection~\cite{wang2021i, zhu2021fooling}. Moreover, some prior works also use it to cause camera effects such as lens flare~\cite{man2020ghostimage} and rolling shutter effect~\cite{kohler2021they} in order to fool the object detection.
    \item \textit{Acoustic signal} has been shown to disrupt or control the outputs of Inertial Measurement Units (IMUs)~\cite{son2015rocking, trippel2017walnut}. Prior work~\cite{ji2021poltergeist} used this attack vector to attack the camera stabilization system, which has built-in IMUs, to manipulate the camera object detection results.
    \item \textit{Translucent patch} refer to sticking translucent films with color spots on camera lens. It can cause misdetect road objects such as vehicles and STOP signs~\cite{zolfi2021translucent}. 
\vspace{-\topsep}
\end{itemize}

\textbf{Cyber-layer attack vectors:}

\begin{itemize}
\vspace{-\topsep}
    \item \textit{ML backdoor} is an attack that tampers the training data or training algorithm to allow the model to produce desired outputs when specific triggers are present. Prior work~\cite{liu2018trojaning} leverages this to attack the end-to-end driving model by presenting the trigger on a roadside billboard.
    \item \textit{Malware \& software compromise} is generic cyber-layer attack vectors assumed in prior works to eavesdrop or modify sensor data~\cite{jha2020ml}, or execute malicious programs alongside the AI components~\cite{luo2020stealthy, hong2020avguardian}.
    Particularly, a domain-specific instance is the Robot Operating System (ROS) node compromises~\cite{hong2020avguardian}, which can modify inter-component messages within ROS-based AD systems, e.g., Apollo v3.0 and earlier~\cite{github_apollo} and Autoware~\cite{github_autoware}.
\vspace{-\topsep}
\end{itemize}

\textit{Status and trends.} 
As shown in Table~\ref{tab:attacks}, existing works predominantly adopt physical-layer attack vectors, with 63.0\% and 27.8\% using physical-world and sensor attack vectors, respectively. Among the physical-world ones, object texture is the most popular (half of the all attacks). This is likely because such attack vectors are direct physical-world adaptations of the digital-space perturbations in general adversarial attacks. In contrast, only 6 (11.1\%) attacks leverage cyber-layer attack vectors. More discussions on this are in \S\ref{sec:gaps_cyber_attack_vectors}.

\subsubsection{Attacker's Knowledge} \label{sec:sok_attack_knowledge} We follow the general definitions of attacker's knowledge by Abdullah et al.~\cite{abdullah2021sok}:

\textbf{White-box.} This setting assumes that the attacker has complete knowledge of the AD system, including the design and implementation of the targeted AI components, corresponding sensor parameters, etc. Among existing attacks, such white-box setting is the most commonly-adopted (Table~\ref{tab:attacks}).

\textbf{Gray-box.} This setting assumes that part of the information required by white-box attacks is unavailable. Prior gray-box attacks all assume the lack of knowledge of AI model internals such as weights. However, some works still require confidence scores in the model outputs~\cite{zhang2019camou, hu2020cca, hamdi2020sada, ji2021poltergeist, wang2021dual, jing2021too, zhu2021can, yang2021robust}, and some require detailed sensor parameters~\cite{wang2021i, sun2021control, ji2021poltergeist}.

\textbf{Black-box.} This is the most restrictive setting where the attacker cannot access any of the internals in the AD vehicle. Prior works that belong to this category are either transfer-based attacks~\cite{lovisotto2021slap, wang2021daedalus}, which generate attack inputs based on local white-box models, or the ones that do not require any model-level knowledge in attack generation~\cite{sun2020towards, kohler2021they}.

\textit{Status and trends.} As shown in Table~\ref{tab:attacks}, existing works commonly assume the white-box settings in their attack designs (66.7\%). However, in recent years, we start to see increasing research efforts in the more challenging but also more practical attack settings, i.e., gray-box (24.1\%) and black-box (9.3\%), mostly published in recent two years (except one). This greatly benefited the practicality and realism of this research space, and also shows that the community practices have evolved significantly from earlier years~\cite{deng2021deep, qayyum2020securing}.

\vspace{-0.1cm}
\nsubsection{Systematization of AD AI Defenses} \label{sec:sok_defenses}

Table~\ref{tab:defenses} summarizes the defenses included in our SoK. We taxonomize them from 4 research aspects critical for the security field: Defense methods, defense goals, defense deployability, and robustness to adaptive attacks.

\begin{table*}[]
\centering
\setlength{\tabcolsep}{2.5pt}
\begin{tabular}{|c?c|r|cc|l|c|cccccccccc|cc|cc|c|}
\cline{1-22}
  \multicolumn{2}{|c|}{} &
  \multicolumn{1}{c|}{} &
  \multicolumn{1}{c}{} &
  \multicolumn{1}{c|}{} &
  \multicolumn{1}{c|}{} &
  \multicolumn{1}{c|}{} &
  \multicolumn{10}{c|}{Deployability} &
  \multicolumn{2}{c|}{} &
  \multicolumn{2}{c|}{\begin{tabular}[c]{@{}c@{}}Eval.\\ level\end{tabular}} &
  \multicolumn{1}{c|}{}
  \\
  \multicolumn{2}{|c|}{Target AI Component} &
  \multicolumn{1}{c|}{Paper} &
  \multicolumn{1}{c}{\vthead{Year}} &
  \multicolumn{1}{c|}{\vthead{Field}} &
  \multicolumn{1}{c|}{Defense method} &
  \vthead{Defense goal} &
  \multicolumn{2}{c}{\vthead{Neg. timing overhead}} &
  \multicolumn{2}{c}{\vthead{Neg. resource overhead}} &
  \multicolumn{2}{c}{\vthead{No model training}} &
  \multicolumn{2}{c}{\vthead{No additional dataset}} &
  \multicolumn{2}{c|}{\vthead{No h/w modification}} &
  \multicolumn{2}{c|}{\vthead{Robust to adaptive attack}} &
  \multicolumn{1}{c}{\vthead{Component-level}} &
  \multicolumn{1}{c|}{\vthead{System-level}} &
  \multicolumn{1}{c|}{\vthead{Open source}}
  \\ \hline
  & 
  & 
  Nassi et al.~\cite{nassi2020phantom} &  
  '20 & 
  S & 
  Driving context \& physics consistency checking & 
  D & 
  \multicolumn{2}{c}{\cmark} & 
  \multicolumn{2}{c}{} & 
  \multicolumn{2}{c}{} & 
  \multicolumn{2}{c}{} & 
  \multicolumn{2}{c|}{\cmark} & 
  \multicolumn{2}{c|}{\cmark} & 
  \cmark & 
   & 
  \cmark 
  \\
  & 
  Object & 
  Li et al.~\cite{li2020connecting} &  
  '20 &  
  V &  
  Driving context consistency checking & 
  D & 
  \multicolumn{2}{c}{\unsure} & 
  \multicolumn{2}{c}{} & 
  \multicolumn{2}{c}{} & 
  \multicolumn{2}{c}{} & 
  \multicolumn{2}{c|}{\cmark} & 
  \multicolumn{2}{c|}{\unsure} & 
  \cmark & 
   & 
  \cmark 
  \\
  Camera & 
  detection & 
  Liu et al.~\cite{liu2021seeing} &  
  '21 &  
  S &  
  Sensor fusion based consistency checking & 
  D & 
  \multicolumn{2}{c}{} & 
  \multicolumn{2}{c}{} & 
  \multicolumn{2}{c}{} & 
  \multicolumn{2}{c}{} & 
  \multicolumn{2}{c|}{} & 
  \multicolumn{2}{c|}{\cmark} & 
  \cmark & 
   & 
  \\
  perception & 
  & 
  Chen et al.~\cite{chen2021class} &  
  '21 &  
  V &  
  Adversarial training & 
  M & 
  \multicolumn{2}{c}{\cmark} & 
  \multicolumn{2}{c}{\cmark} & 
  \multicolumn{2}{c}{} & 
  \multicolumn{2}{c}{\cmark} & 
  \multicolumn{2}{c|}{\cmark} & 
  \multicolumn{2}{c|}{\unsure} & 
  \cmark & 
   & 
  \\ \cline{2-22}
  & 
  Obj. tracking & 
  Jia et al.~\cite{jia2020robust} & 
  '20 &  
  V &  
  Predict \& remove perturbation from inputs & 
  M & 
  \multicolumn{2}{c}{\unsure} & 
  \multicolumn{2}{c}{} & 
  \multicolumn{2}{c}{} & 
  \multicolumn{2}{c}{} & 
  \multicolumn{2}{c|}{\cmark} & 
  \multicolumn{2}{c|}{\unsure} & 
  \cmark & 
  & 
  \cmark 
  \\ \hline
  \multicolumn{2}{|c|}{Camera perception \& localization} & 
  Wang et al.~\cite{wang2021i} &  
  '21 &  
  S & 
  Consistency checking on reflected lights & 
  D & 
  \multicolumn{2}{c}{\unsure} & 
  \multicolumn{2}{c}{} & 
  \multicolumn{2}{c}{} & 
  \multicolumn{2}{c}{} & 
  \multicolumn{2}{c|}{\cmark} & 
  \multicolumn{2}{c|}{\unsure} & 
  \cmark & 
   & 
  \\ \hline
  & 
  & 
  Sun et al.~\cite{sun2020towards} & 
  '20 &  
  S &  
  Consistency checking on point cloud free space & 
  D & 
  \multicolumn{2}{c}{\cmark} & 
  \multicolumn{2}{c}{\cmark} & 
  \multicolumn{2}{c}{\cmark} & 
  \multicolumn{2}{c}{\cmark} & 
  \multicolumn{2}{c|}{\cmark} & 
  \multicolumn{2}{c|}{\cmark} & 
  \cmark & 
  & 
  \\ 
  LiDAR perception & 
  Obj. detection & 
  Sun et al.~\cite{sun2020towards} & 
  '20 &  
  S &  
  Augment inputs with obj. conf. from front view & 
  M & 
  \multicolumn{2}{c}{\unsure} & 
  \multicolumn{2}{c}{} & 
  \multicolumn{2}{c}{} & 
  \multicolumn{2}{c}{\cmark} & 
  \multicolumn{2}{c|}{\cmark} & 
  \multicolumn{2}{c|}{\cmark} & 
  \cmark & 
  & 
  \\ 
  & 
  & 
  You et al.~\cite{you2021temporal} & 
  '21 &  
  S &  
  Trajectory consistency checking & 
  D & 
  \multicolumn{2}{c}{\cmark} & 
  \multicolumn{2}{c}{} & 
  \multicolumn{2}{c}{\cmark} & 
  \multicolumn{2}{c}{\cmark} & 
  \multicolumn{2}{c|}{\cmark} & 
  \multicolumn{2}{c|}{\unsure} & 
  \cmark & 
  & 
  \\ \hline
\end{tabular}
\vspace{0.03in}

\footnotesize
Field: S = Security, V = Computer Vision, M = ML/AI, O = Others, e.g., Robotics, arXiv;
Defense goal: D = detection, M = mitigation,\\
\unsure~= not available in the paper and we cannot conclude from the design, conf. = confidence, h/w = hardware

\vspace{-0.2cm}
\caption[overview]{Summary of existing semantic AD AI defense solutions in our SoK scope (\S\ref{sec:background_scope}).}
\label{tab:defenses}
\vspace{-0.27in}
\end{table*}

\subsubsection{Defense Methods} \label{sec:sok_defense_methods} 
In general, the defense methods in existing works can be categorized into two categories:

\textbf{Consistency checking.}
Consistency checking is a general attack detection technique that cross-checks the attacked information either with other (ideally independent) measurement sources, or with invariant properties of itself. For example, prior works use the detected objects from stereo cameras~\cite{liu2021seeing} and object trajectory from prediction models~\cite{you2021temporal} to cross-check LiDAR object detection results. Li et al.~\cite{li2020connecting} and Nassi et al.~\cite{nassi2020phantom} created detection models to determine whether the current camera object detection results are consistent with the driving context. Some works also leverage physical invariant properties of the sensor source such as light reflection~\cite{nassi2020phantom, wang2021i} and LiDAR occlusion patterns~\cite{sun2020towards} to detect AI attacks.

\textbf{Adversarial robustness improvement.} Several defenses try to improve the robustness of the AI component against attacks. For example, Chen et al.~\cite{chen2021class} applied adversarial training~\cite{goodfellow2014explaining} to make the camera object detection model more robust. Jia et al.~\cite{jia2020robust} improved the model robustness by predicting and removing potential adversarial perturbations from the model inputs. Sun et al.~\cite{sun2020towards} augment the point cloud with point-wise confidence scores from a front view segmentation model to improve the robustness of LiDAR object detection.

\textit{Status and trends.} As shown in Table~\ref{tab:defenses}, existing defenses share common general defense strategies, with 66.7\% (6/9) based on consistency checking and 33.3\% (3/9) based on adversarial robustness improvement. We discuss a few possible new directions later in~\S\ref{sec:gaps_lack_of_defense}.

\subsubsection{Defense Goals} \label{sec:sok_defense_goals} 
The defense methodologies in the above section can be naturally mapped to two defense goals: (1) \textbf{Detection:} All consistency checking based defenses are designed to detect the attack attempts; and (2) \textbf{Mitigation:} Improving the adversarial robustness of models can generally reduce the attack success rate and raise the bar of the attackers. However, it is not designed to detect the attack, and also cannot fundamentally eliminate/prevent AI vulnerabilities. 

\textit{Status and trends.} Among the existing defenses, attack detection and mitigation are the main focus; so far, none of the works aims for other goals such as attack prevention. More discussions on this are in~\S\ref{sec:gaps_lack_of_defense}.

\subsubsection{Defense Deployability} \label{sec:sok_defense_depployability}
In this section, we list the defense design properties that are highly desired for the deployment in practical AD settings:

\textbf{Negligible timing overhead.} Timing overhead is one of the most important factors that may limit the deployability of defense for real-time systems such as AD systems. Among the existing defenses, 4 have negligible or no timing overhead by design or shown in evaluation~\cite{nassi2020phantom, chen2021class, sun2020towards, you2021temporal}, 1 fails to catch up with the camera and LiDAR frame rate in evaluation~\cite{liu2021seeing}. For the remaining ones, we cannot conclude their timeliness from their papers (e.g., no timing overhead evaluation).

\textbf{Negligible resource overhead.} In practice, the hardware that runs the AD system may have limited computation resources (e.g., limited GPUs) due to budget concerns. Among the defenses, except the ones that do not require \textit{extra} ML models~\cite{chen2021class, sun2020towards, sun2021control}, all others will impose additional demand on the computation resources in order to execute the models. This may prevent them from deploying onto vehicles that have already fully utilized the hardware resources.

\textbf{No model training.} The requirement of model training imposes extra deployment burdens. Among the defenses that require extra ML models, only one~\cite{you2021temporal} uses public pre-trained models without fine-tuning.

\textbf{No additional dataset.} Closely related to the previous property, some defenses not only require model training, but also need additional datasets during training process. This will limit the deployability due to the efforts in dataset preparation.

\textbf{No hardware modification.} This property means whether the defense requires changes to the hardware on AD vehicle or adding new hardware (e.g., additional sensors). Among the defenses, Liu et al.~\cite{liu2021seeing} require stereo cameras, which may not be available on some AD vehicles.

\textit{Status and trends.} As shown in Table~\ref{tab:defenses}, almost all existing defenses (7/8) have the awareness for at least one of these deployability design aspects, especially for that regarding hardware modification (7/8) and timing overhead (4/8). However, the awareness on the need for no model training, negligible resource overhead, and no additional dataset are currently lacking (2/8, 2/8, and 3/8 respectively).

\subsubsection{Robustness to Adaptive Attacks} \label{sec:sok_defense_adaptive_attacks} Adaptive attacks are designed to circumvent a particular defense. They often assume complete knowledge of the defense internals, including the design and implementation, and are designed to challenge the fundamental assumptions of the defense. Adaptive attack evaluation has become \textit{strongly-advocated} in recent adversarial AI defenses, and prior work has also proposed guidelines on how to properly design adaptive attacks~\cite{tramer2020adaptive}. However, among the AD AI defenses, we find that \textit{only 3 papers} conduct some forms of adaptive attack evaluation~\cite{nassi2020phantom, liu2021seeing, sun2020towards}. Specifically, Nassi et al.~\cite{nassi2020phantom} applied existing adversarial attacks on their defense model and find that it is challenging to circumvent the attack detection; Liu et al.~\cite{liu2021seeing} evaluated simultaneous attacks against their consistency checking design between LiDAR and camera object detection, and find that the detection is still effective. Sun et al.~\cite{sun2020towards} designed adaptive attacks based on the defense assumptions (the LiDAR occlusion patterns), and show that they fail to break the two defenses.

\textit{Status and trends.} Despite being strongly-advocated in general adversarial AI defense research~\cite{tramer2020adaptive}, currently not many (3/8) of the existing defenses in AD AI security conduct evaluation against adaptive attacks. However, with the increasing adoption of such practices in the general adversarial AI domain, this situation on the semantic AD AI security domain should also see improvements.

\vspace{-0.1cm}
\nsubsection{Systematization of Evaluation Methodology} \label{sec:sok_eval_method}

At the evaluation methodology level, besides the expected differences due to problem formulations (e.g., attack targets and goals), we notice an interesting general disparity in the choices of the \textit{evaluation levels}, which is relatively unique for semantic AI security as opposed to generic AI security. Specifically, since here the attack targets are AI components but the ultimate goals are to achieve system-level attack effect (e.g., crashes), the evaluation methodology can be designed at both AI component and AD system levels:

\textbf{Component-level evaluation} is to evaluate attack/defense impacts only at the targeted AI component level (e.g., object detection) without involving (1) other necessary components in the full-stack AD systems (e.g., object tracking, prediction, planning); and (2) the closed-loop control~\cite{franklin2002feedback}. The evaluation metrics are thus component-specific, e.g., detection rate~\cite{eykholt2018physical, zhao2019seeing}, location deviation~\cite{shen2020drift}, steering error~\cite{kong2020physgan}, etc. 

\textbf{System-level evaluation} refers to evaluating the attack/defense impacts at the vehicle driving behavior level considering full-stack AD system pipelines and closed-loop control. The evaluation setups used to achieve this can be generally classified into two categories: (1) \textit{Real vehicle-based}, where a real vehicle is under partial (e.g., steering only~\cite{jing2021too}) or full (i.e., steering, throttle, and braking~\cite{nassi2020phantom, sun2021control}) closed-loop control of the AD system to perform certain driving maneuvers with the presence of attacks, which are typically deployed in the physical world as well; and (2) \textit{Simulation-based}, where the critical elements in the closed-loop control (e.g., sensing/actuation hardware and the physical world) are fully or partially simulated, using either existing simulation software~\cite{sato2021dirty, tang2021fooling, cao2021invisible, shen2020drift, liu2018trojaning} or custom-built modeling~\cite{shen2020drift}. 

\textit{Status and trends}. As shown in Table~\ref{tab:attacks} and~\ref{tab:defenses}, the majority (90.5\%) of the surveyed works performed component-level evaluation, likely due to the ease of experiment efforts (no need to set up real vehicle or simulation), while only 25.4\% (16/63) adopted some forms of system-level evaluation. More discussions are in~\S\ref{sec:gaps_system_level_eval}.
\vspace{-0.1cm}
\nsection{Scientific Gaps and Future Directions} \label{sec:gaps}

Based on the systematization of existing works on AD AI security, we summarize a list of scientific gaps that we observed and also discuss possible solution directions. To avoid subjective opinions and bias, such observations are drawn from quantitative comparisons \textit{vertically} among the choices made in existing AD AI security works along with different design angles in~\S\ref{sec:sok} and/or \textit{horizontally} with security works in closely-related CPS (Cyber-Physical Systems) domains such as drone and Automatic Speech Recognition and Speaker Identification (ASR/SI) based on recent SoKs~\cite{nassi2021sok, abdullah2021sok}.

\vspace{-0.1cm}
\nsubsection{Evaluation: General Lack of System-Level Evaluation} \label{sec:gaps_system_level_eval}

\begin{gap} \label{gap:system_level_eval}
It is widely recognized that in AD system, AI component-level errors do not necessarily lead to system-level effect (e.g., vehicle collisions). However, system-level evaluation is generally lacking in existing works.
\end{gap}
\vspace{-0.2cm}
As identified in~\S\ref{sec:sok_eval_method}, currently it is not a common practice for AD AI security works to perform system-level evaluation: overall only 25.4\% of existing works perform that, and such a number is especially low (\textit{7.4\%}) for the most extensively-studied AI component, camera object detection. In fact, the vast majority (74.6\%) of these works \textit{only} performed component-level evaluation without making any efforts to experimentally understand the system-level impact of their attack/defense designs. Admittedly, for CPS systems such system-level evaluation is generally more difficult due to the involvement of physical components. However, we find that for existing security research on related CPS domains such as drone and ASR/SI, actually \textit{almost all} attack works perform system-level evaluations (100\% for drone, 94\% for ASR/SI based on the SoKs~\cite{nassi2021sok, abdullah2021sok}). This shows that the current AD AI security research community is particularly lagging behind in such common evaluation practice in CPS security research.

In the CPS verification area, it is actually already widely-recognized that for CPS with AI components, \textit{AI component-level errors do not necessarily lead to system-level effects}~\cite{dreossi2019compositional, seshia2020semantic}. For AD systems, this is especially true due to the high end-to-end system-level complexity and closed-loop control dynamics, which can explicitly or implicitly create fault-tolerant effects for AI component-level errors. In fact, various such counterexamples have already been discovered in AD system context, e.g., when the object detection model error is at a far distance for automatic emergency braking systems~\cite{dreossi2019compositional, seshia2020semantic}, or such errors can be effectively tolerated by downstream AI modules such as object tracking~\cite{jia2020fooling}. This means that even with high attack success rates shown at the AI component level, it is actually possible that such an attack \textit{cannot cause any meaningful effect to the AD vehicle driving behavior}. For example, as concretely estimated by Jia et al.~\cite{jia2020fooling}, for camera object detection-only AI attacks, a component-level success rate of up to 98\% can still be not enough to affect object tracking results. Thus, we believe that for semantic AD AI security research, the current general lack of system-level evaluation is a critical \textit{scientific methodology-level gap} that should be addressed as soon as possible. 

\textbf{Future directions.} Specific to the AD problem domain, there are various technical challenges to effectively fill this gap at the research community level. Among the possible options (\S\ref{sec:sok_eval_method}), real vehicle-based system-level evaluation is generally unaffordable for most academic research groups (e.g., \$250K per vehicle~\cite{cost_level_4_av}), not to mention the need for easily-accessible testing facilities, the high time and engineering efforts needed to set up the testing environment, and also high safety risks (especially since most AD AI attacks aim at causing safety damages). Simulation-based evaluation is much more affordable, accessible, and safe, but researchers still need to spend substantial engineering efforts to instrument existing AD simulation setup for the security-specific research needs, e.g., modifying the simulated physical environment rendering~\cite{sato2021dirty, cao2021invisible} or sensing channels~\cite{shen2020drift} for launching attacks. These might explain why system-level evaluation is not commonly adopted for AD AI security research today.

To address such impasse, it is highly desired if there can be a community-level effort to collectively build a common system-level evaluation infrastructure, since (1) the engineering efforts spent in instrumenting either a real AD vehicle or AD simulation for security research evaluation share common design/implementation patterns (e.g., common attack entry points as in~\S\ref{sec:sok_attack_vectors}); and (2) in AD context, the system-level attack effect can be highly influenced by driving scenario setups (e.g., large braking distance differences in highway and local roads~\cite{hancock2013policy} and thus the system-level evaluation results are only comparable (and thus scientifically-meaningful) if the same evaluation scenario and metric calculation are used. Considering the criticality of such a methodology-level scientific gap, later in~\S\ref{sec:platform} we take the initiative to lay the groundwork to foster such a community-level effort.

\vspace{-0.1cm}
\nsubsection{Research Goal: General Lack of Defense Solutions} \label{sec:gaps_lack_of_defense}

\begin{gap} \label{gap:lack_of_defense}
Compared to AD AI security attacks, their effective defense solutions are substantially lacking today, especially those on attack prevention.
\end{gap}
\vspace{-0.2cm}

As shown in Tables~\ref{tab:attacks} and~\ref{tab:defenses}, among existing semantic AD AI security works, the ratio of discovered attacks (85.7\%) is much higher than effective defense solutions (14.3\%). Furthermore, all existing defenses focus on attack detection and mitigation, and \textit{none} studied \textit{prevention}, which is a more ideal form of defense. In comparison, such ratios are much more balanced in the drone security field~\cite{nassi2021sok}, where 51\% are on attacks and 49\% are on defenses. Also, among the defenses, \textit{33\%} are for prevention.
While attack discovery is the necessary first step for security research into an emerging area, it is imperative to follow up with effective defense solutions, especially those on attack prevention, to close the loop of the discovered attacks and actually benefit the society.

\textbf{Future directions.} As shown in Tables~\ref{tab:attacks} and~\ref{tab:defenses}, many of the AD AI components with discovered attacks actually do not have any effective defense solutions yet (e.g., lane detection, MSF perception), which are thus concrete defense research directions. Note that there indeed exist generic AI defenses that are directly applicable (e.g., input transformation~\cite{Xu:2018:ndss}), but it is already found that such generic defenses are generally ineffective against the CPS domain AI attacks, e.g., for both AD~\cite{sato2021dirty, cao2021invisible} and ASR/SI~\cite{hussain2021waveguard}.

Besides considering the currently-undefended AI components, there are also possibly-applicable defense strategies that have not been explored yet, for example certified robustness~\cite{lecuyer2019certified}. While unexplored (\S\ref{sec:sok_defense_methods}), such a defense is actually highly desired in the AD AI security domain since it can provide strong theoretical guarantees for AI security in such a safety-critical application domain. The challenge is that today's certified robustness designs only focus on small 2D digital-space perturbations, and thus their extensions to today's popular AD AI attack vectors (e.g., physical-world or sensor attacks) are still open research problems. This requires addressing at least the system-to-AI semantic gap, not to mention potential deployability challenges as such methods generally have high overhead~\cite{xiang2021detectorguard} and thus might be hard to meet the strong real-time requirement in AD context (\S\ref{sec:sok_defense_depployability}).

\vspace{-0.1cm}
\nsubsection{Attack Vector: Cyber-Layer Attack Vectors Under-Explored} \label{sec:gaps_cyber_attack_vectors}

\begin{gap} \label{gap:cyber_attack_vectors}
Existing works predominately focus on physical-layer attack vectors, which leaves the cyber-layer ones substantially under-explored.
\end{gap}
\vspace{-0.2cm}
As shown in Table~\ref{tab:attacks}, there are only 11.1\% (6/54) AD AI attack works that leverage cyber-layer attack vectors in their attack designs, assuming ML backdoors~\cite{liu2018trojaning}, malware~\cite{jha2020ml}, remote exploitation~\cite{luo2020stealthy}, and compromised ROS nodes~\cite{hong2020avguardian}, respectively. Among these four, only the last one is relatively domain-specific to AD. In related CPS domains such as drones and ASR/SI, such ratio is much higher ($>$50\%): 53.8\% for drone attack works~\cite{nassi2021sok} and 52.9\% for ASR/SI ones~\cite{abdullah2021sok}. One observation we have is that many of these works in related CPS domains exploit domain-specific networking channels such as ground station communication channel~\cite{luo2016drones} and First-Person View (FPV) channel~\cite{rodday2016hacking,multerer2017low, deligne2012ardrone} in drones, audio files, and telephone networks~\cite{abdullah2021hear} in ASR/SI systems. However, \textit{no existing works in AD AI security} studied such aspects, which can thus be one direction to fill this research gap.

\textbf{Future directions.}  The general design goal of AD AI stack is indeed mostly towards achieving autonomy only using on-board sensing; however, this does not mean that there are no networking channels that can severely impact end-to-end driving. For example, at least the following two are domain-specific, widely-adopted in real-world AD vehicles, and highly security and safety critical:

\underline{\textit{High Definition (HD) Map update channel.}} For L4 vehicles, the accuracy of HD Map is critical for driving safety as it is the fundamental pillar for almost all AD modules such as localization, prediction, routing, and planning. Real world dynamic factors (e.g., road constructions), make it imperative to keep the HD Map frequently updated, which is typically over-the-air. For example, when a Waymo AD vehicle detects a road construction, it updates the HD Map and shares the map update with the operations center and the rest of the fleet in real time~\cite{waymo_hdmap_update}.

\underline{\textit{Remote AD operator control channel.}}
Another channel that is also relevant to high-level AD is the control channel for remote operators. Unlike lower-level AD vehicles, L4 and above do not have safety drivers onboard. However, after an AD vehicle reaches a fail-safe state, a remote operator usually take over the control (e.g., happened to a Waymo vehicle stuck on the road~\cite{waymo_remote_operator}), which is directly safety-critical if hijacked.

\vspace{-0.1cm}
\nsubsection{Attack Target: Downstream AI Component Under-Explored} \label{sec:gaps_downstream_components}

\begin{gap} \label{gap:downstream_components}
There is a substantial lack of works focusing on downstream AI components (e.g., prediction, planning), which are equally (if not more) important compared to upstream ones w.r.t system-level attack effect.
\end{gap}
Among the vast majority (50/54) of attack works that target the more practical modular AD system designs (\S\ref{sec:background_ad_ai}), so far \textit{none} of them targeted downstream AI components such as prediction and planning; they predominately focus on the upstream ones such as perception and localization. This is understandable since under the most popularly-considered physical-layer attack model (\S\ref{sec:sok_attack_vectors}), upstream ones can be much more directly influenced by attack inputs (e.g., via physical-world or sensor attacks), while to affect the inputs of downstream ones such as predication, one has to manipulate the upstream component outputs (e.g., object detection) first. However, these downstream ones are actually at least equally (if not arguably more) important than the upstream ones. For example, errors in the obstacle trajectory prediction or path planning will actually more directly affect driving decisions and thus lead to system-level effects. Thus, it is highly desired for future AD AI security research to fill this gap.

\textbf{Future directions.} To study downstream AI component security, a general challenge is how to study their component-specific security properties with sufficient isolation with upstream component vulnerabilities, so that we can isolate the causes and gain more security design insights specific to the targeted downstream component. Here we discuss several possible solution directions:

\underline{\textit{Physical-layer attacks by road object manipulation.}} The prediction component in AD systems predicts obstacle trajectory based on its detected physical properties (e.g., obstacle type, dimension, position, orientation, speed). Therefore, assuming upstream components such as AD perception is functioning correctly, the attacker can directly control a road object in the physical world (e.g., an attack vehicle on the road), in order to control the inputs of prediction without relying on vulnerabilities in upstream components. However, this requires the attack design to ensure that the triggers of the prediction vulnerability are \textit{semantically-realizable}, e.g., the attacker-controlled obstacle needs to have a consistent type, no breaks in the historical trajectory, moving at a reasonable speed, etc.

\underline{\textit{Physical-layer attacks by localization manipulation.}} The planning component takes the prediction and localization outputs as inputs to calculate the optimal driving path in the near future. Therefore, the change in the localization directly affects the decision-making in the planning. To attack the planning, one can leverage localization-related attack vectors (e.g., GPS spoofing) to control planning inputs. However, this only works for AD systems that purely rely on such input (e.g., GPS) for localization; if MSF-based localization is used, it is still hard to isolate planning-specific vulnerabilities from MSF algorithm vulnerabilities (e.g., \cite{shen2020drift}).

\underline{\textit{Cyber-layer attacks.}} Perhaps the most direct way to manipulate the inputs to the downstream components is via cyber-layer attacks. For example, a compromised ROS node~\cite{hong2020avguardian} in the AD system can directly send malicious messages to prediction or planning. Unlike the road object manipulation method, this does not require the inputs to be semantically-realizable. This thus forms another motivation to explore more on cyber-layer attacks (\S\ref{sec:gaps_cyber_attack_vectors}).

\vspace{-0.1cm}
\nsubsection{Attack Goal: Goals Other Than Integrity Under-Explored} \label{sec:gaps_attack_goals}

\begin{gap} \label{gap:attack_goals}
Existing attacks predominantly focus on safety/traffic rule violations or mobility degradation (can be viewed as integrity in AD domain), while other important security properties such as confidentiality/privacy, availability, and authenticity are under-explored.
\end{gap}
\vspace{-0.2cm}

As shown in~\S\ref{sec:sok_attack_goals}, almost all (96.3\%) existing attacks target the AD-specific manifestation of integrity (i.e., modify the driving correctness). However, the study of other important security properties such as confidentiality and availability~\cite{stallings2012computer} is substantially lacking: only 2 (3.7\%) touch upon confidentiality and none of them study availability. In comparison, the attack goals in drone security are more balanced: 57.9\% (11/19) on integrity/safety, 31.6\% (6/19) on privacy/confidentiality, and 88.9\% (8/19) on availability. Although integrity is highly important in AD context as vehicles are heavy and fast-moving, these other properties are also equally important from the general security field perspective.

\textbf{Future directions.} To foster future research into these less-explored security properties, here we discuss a few possible new research angles. For confidentiality/privacy, existing efforts only consider AD system-internal information leakage  (e.g., location), but one can also consider potential private information extraction from AD system-\textit{external} information such as from sensor inputs. In the drone security area, one main privacy attack scenario is along this direction, e.g., people-spying using the drone camera~\cite{wang2016flying, crypta2019nassi}. For availability, since it is closely related to the communication channels among AI components, more extensive exploration on the cyber-layer attacks (\S\ref{sec:gaps_cyber_attack_vectors}) may make such attacks feasible. So far, no existing works consider authenticity in AD context, but in today's AD deployment and commercialization, authenticity can manifest as the authentication of the safety drivers, mutual authentication between the passenger and the AD vehicle for robotaxi, and mutual authentication between the consumers and the delivery vehicle for AD goods delivery, which are within the broader scope of AD AI stack.

\vspace{-0.1cm}
\nsubsection{Community: Substantial Lack of Open Sourcing} \label{sec:gaps_open_source}

\begin{gap} \label{gap:open_source}
The open-sourcing status of AD AI security works from the security community is much lacking compared to related communities/domains (e.g., CV, ML/AI, ASR/SI), which harms the reproducibility and comparability for this emerging area in the security community.
\end{gap}
\vspace{-0.2cm}
For each work in Table~\ref{tab:attacks} and Table~\ref{tab:defenses}, we searched for their code or open implementation in the paper and also on the Internet. Overall, there are less than \textit{20.6\% (7/34)} papers from security conferences that release source code. The situation is much worse if we narrow it down to the sensor attack works, where only \textit{1 (8.3\%) out of 12} works released its attack code. In comparison, the papers published in CV and ML/AI conferences have over 50\% open-source percentages. Interestingly, also in the CPS domain, 50\% ASR/SI papers in the security conferences release their code, which is roughly the same as in CV and ML/AI conferences. This indicates that it is not that security researchers are not willing to share code, but more likely related to the \textit{diversity} of AD AI security papers in the security conferences. In fact, security conference papers adopt a much more diverse set of attack vectors than any other conferences as shown in Table~\ref{tab:attacks}. For example, among the 15 works that use sensor attack vectors, only 2 are from CV conferences and 1 is from ML/AI conferences. For those papers, it is indeed more difficult to share the code or implementation since hardware designs are usually involved. In comparison, the attack vectors in ASR/SI are much more uniform; they predominantly leverage malicious sound waves, which is convenient to modify, and can be evaluated digitally.

\textbf{Future directions.} There is no doubt that we should encourage more open-sourcing efforts from the security community such that future works can benefit from it. However, the challenge is in which form the hardware implementations should be shared such that such sharing can be more directly useful for the community. Here we discuss two possible solution directions based on our observations from prior works:

\underline{\textit{Open-source hardware implementation references.}} Since the reproduction cost and effort for hardware implementations are generally higher than software-only ones, it is actually more desired for researchers to release as many details about their hardware design as possible. This often means the circuit diagram, printed circuit board layout, bill of materials, and detailed experiment procedures. One example is the LiDAR spoofing work by Shin et al.~\cite{shin2017illusion}, where they clearly list such details on a website and thus other groups were able to reproduce and build upon their designs~\cite{cao2019adversarial}.

\underline{\textit{Open-source attack modeling code.}} Another interesting trend in recent sensor attack works is that they often model the attack capability in the digital space for large-scale evaluation. For example, Man et al.~\cite{man2020ghostimage} modeled the camera lens flare effect caused by attacker's light beams in digital images, Ji et al.~\cite{ji2021poltergeist} modeled the image blurry effect caused by adversarial acoustic signals on the camera stabilization system; Cao et al.~\cite{cao2019adversarial} models the LiDAR spoofing capability as the number of points that can be injected to the point cloud; Shen et al.~\cite{shen2020drift} models GPS spoofing as arbitrary GPS position that can be controlled by the attacker. Since these modelings are either validated in the physical-world experiments or backed up by the literature, sharing such code will greatly ease the reproduction in future works.

Our initiated community-level evaluation infrastructure development effort in~\S\ref{sec:platform} may also help fill this gap by encouraging open-sourcing practices (\S\ref{sec:platform_design_goal_choices}). 
\vspace{-0.1cm}
\nsection{\textit{PASS}: Uniform and Extensible System-Driven Evaluation Platform for AD AI Security} \label{sec:platform}

Among all the identified scientific gaps, the one on the general lack of system-level evaluation (\S\ref{sec:gaps_system_level_eval}) is especially critical as proper evaluation methodology is crucial for valid scientific progress. In this section, we take the initiative to address this critical scientific methodology-level gap by developing an open-source, uniform, and extensible system-driven evaluation platform, named \textit{PASS} (\underline{P}latform for \underline{A}utonomous driving \underline{S}afety and \underline{S}ecurity).

\vspace{-0.1cm}
\nsubsection{Design Goals and Choices}
\label{sec:platform_design_goal_choices}

As described in~\S\ref{sec:gaps_system_level_eval}, to effectively fill this gap at the research community level, a common system-level evaluation infrastructure is highly desired. To foster this, we take the initiative to build a \textit{system-driven evaluation platform} that unifies the common system-level instrumentations needed for AD AI security evaluation based on our SoK (\S\ref{sec:sok}), and abstracts out the customized attack/defense implementation and experimentation needs as platform APIs. Such a platform thus directly provides the aforementioned evaluation infrastructure as it allows semantic AD AI security works to directly plug in their attack/defense designs to obtain system-level evaluation results, such that (1) without the need to spend time and efforts to build the lower-level evaluation infrastructure; and (2) generating results in the same evaluation environment and thus directly comparable to prior works. Note that although some existing works developed simulation-based system-level testing methods for AI-based AD~\cite{dreossi2019compositional, o2017computer, tuncali2018simulation, dreossi2019verifai}, their designs are for safety not security (no threat models), and thus cannot readily support the evaluation of different AD AI attack/defense methods.

This platform will be fully open-sourced so that researchers can collectively develop new APIs to fit future attack/defense evaluation needs, and also contribute attack and defense implementations to form a semantic AD AI security benchmark, which can improve comparability, reproducibility, and also encourage open-sourcing (currently lacking in the security community as in~\S\ref{sec:gaps_open_source}). We also plan to provide remote evaluation services for research groups that do not have the hardware resources required to run this platform, which will be maintained by the project team. 

\begin{table}[t]
\centering
\setlength{\tabcolsep}{2.5pt}
\begin{tabular}{|l|cccccc|l|}
\hline
  Method & FD & AF & SF & FX & EF & RP & ~Papers \\ \hline
  Real vehicle-based & 
  \multirow{1}{*}{\pie{4}} & 
  \multirow{1}{*}{\pie{0}} & 
  \multirow{1}{*}{\pie{0}} & 
  \multirow{1}{*}{\pie{0}} & 
  \multirow{1}{*}{\pie{0}} & 
  \multirow{1}{*}{\pie{0}} & 
  \multirow{1}{*}{\cite{nassi2020phantom, jing2021too, sun2021control}} \\
  \hline
  Simulation-based & 
  \multirow{1}{*}{\pie{2}} & 
  \multirow{1}{*}{\pie{4}} & 
  \multirow{1}{*}{\pie{4}} & 
  \multirow{1}{*}{\pie{4}} & 
  \multirow{1}{*}{\pie{4}} & 
  \multirow{1}{*}{\pie{4}} &
  \multirow{1}{*}{\cite{jha2020ml, sato2021dirty, tang2021fooling, yang2021robust, cao2021invisible, luo2020stealthy, shen2020drift, liu2018trojaning}} \\ \hline
\end{tabular}

\vspace{0.05in}
\footnotesize
FD = fidelity, AF = affordability, SF = safety, FX = flexibility, \\ EF = efficiency, RP = reproducibility;
\piel{0} = low, \piel{2} = medium, \piel{4} = high

\caption[overview]{System-level evaluation methods used in existing works. Such complementariness motivates us to choose a \textit{simulation-centric hybrid design} for our evaluation platform.
}
\label{tab:system-level-eval}
\vspace{-0.1in}
\end{table}

\textbf{Simulation-centric hybrid design.} To build such a community-level evaluation infrastructure, a fundamental design challenge is the trade-off between real vehicle-based and simulation-based evaluation methodology, which is shown in~Table~\ref{tab:system-level-eval}. Real vehicle-based is more fidel as it has the vehicle, sensors, and physical environment all in the evaluation loop, but simulation is better in all other important research evaluation aspects, ranging from cost, free of safety issues, high flexibility in scenario customization, convenience of attack deployment, much faster evaluation iterations, to high reproducibility. Considering their strengths and weakness, we choose a \textit{simulation-centric hybrid design}, in which we mainly design for simulation-based evaluation infrastructure and only use real vehicles for simulation fidelity improvement (e.g., digital twin~\cite{svl_digital_twin} and sensor model calibration~\cite{svl_digital_twin, manivasagam2020lidarsim, nakashima2021learning, weston2021there, chen2021geosim}) and exceptional cases where it is easy to set up the physical scenario and maintain safety.

We choose the design to be simulation-centric because we find that the fidelity drawback of simulation-based approach is tolerable since (1) our results show that for today's representative AD AI attacks, the attack results and characteristics are highly correlated in today's industry-grade simulator and physical world, which indicates sufficient fidelity at least for such AD AI security evaluation purpose (detailed in \S\ref{sec:platform}); and (2) the simulation fidelity technology is still evolving as this is also the need for the entire AD industry~\cite{cruise_safety_report, waymo_safety_report}, for example recently there are various new advances in both industry~\cite{svl_digital_twin} and academia~\cite{svl_digital_twin, manivasagam2020lidarsim, nakashima2021learning, weston2021there, chen2021geosim}); and (3) after all the simulation environment can be considered as a different environmental domain, and practical attacks/defenses should be capable of such domain adaptations. Such a design is also more beneficial from the community perspective as it makes the setup of such a platform more affordable for more research groups.

\vspace{-0.1cm}
\nsubsection{PASS Design} \label{sec:platform_design}

Our system-driven evaluation platform \textit{PASS} is guided by two general design rationales: (1) \textit{uniform}, aiming at unifying the common system-level evaluation needs at attack/defense implementation, scenario setup, and evaluation metric levels observed in our SoK (\S\ref{sec:sok}); and (2) \textit{extensible}, aiming at making the platform capable of conveniently supporting new future attacks/defenses, AD system designs, and evaluation setups. Specifically, for attacks/defenses on a certain driving task, our platform defines a list of driving scenarios and metrics, which help unify the experimental settings and effectiveness measurements. To enable extensibility for future research, the platform provides a set of interfaces to simplify the deployment of new attacks/defenses. Based on our SoK, we design the attack/defense interfaces to support the common categories of attack vectors (\S\ref{sec:sok_attack_vectors}) and defense strategies (\S\ref{sec:sok_defense_methods}). In addition, the platform provides a modular AD system pipeline with pluggable AI components, which makes the platform easily extensible to different AD system designs.

\begin{figure}[tbp]
\centering
\vspace{-0.15cm}
\includegraphics[width=\linewidth]{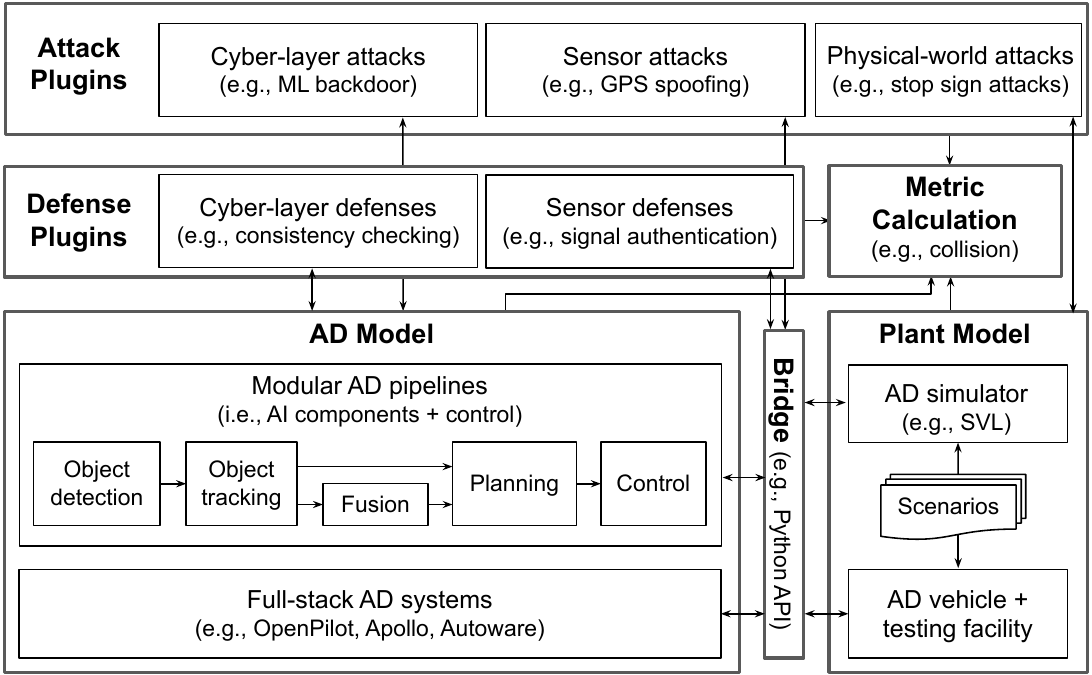}
\caption{Design of our system-driven evaluation platform \textit{PASS} for the semantic AD AI security community.}
\label{fig:platform}
\vspace{-0.3cm}
\end{figure}

Fig.~\ref{fig:platform} shows the \textit{PASS} design, consists of 6 main modules:

\subsubsection{AD model} The AD model hosts the end-to-end AD system with the targeted AI components under testing. Specifically, the platform provides two AD model choices: modular AD pipeline and full-stack AD system (e.g., OpenPilot~\cite{news_openpilot}, Apollo~\cite{github_apollo}, and Autoware~\cite{github_autoware}). The modular AD pipeline is provided as a flexible option to enable AD systems with replaceable AI components and different system structures. To configure the modular AD pipeline, the user only needs to modify a human-readable configuration file to specify the desired AI components to be included.

\subsubsection{Plant model} 
\begin{figure}[tbp]
\centering
\includegraphics[width=\linewidth]{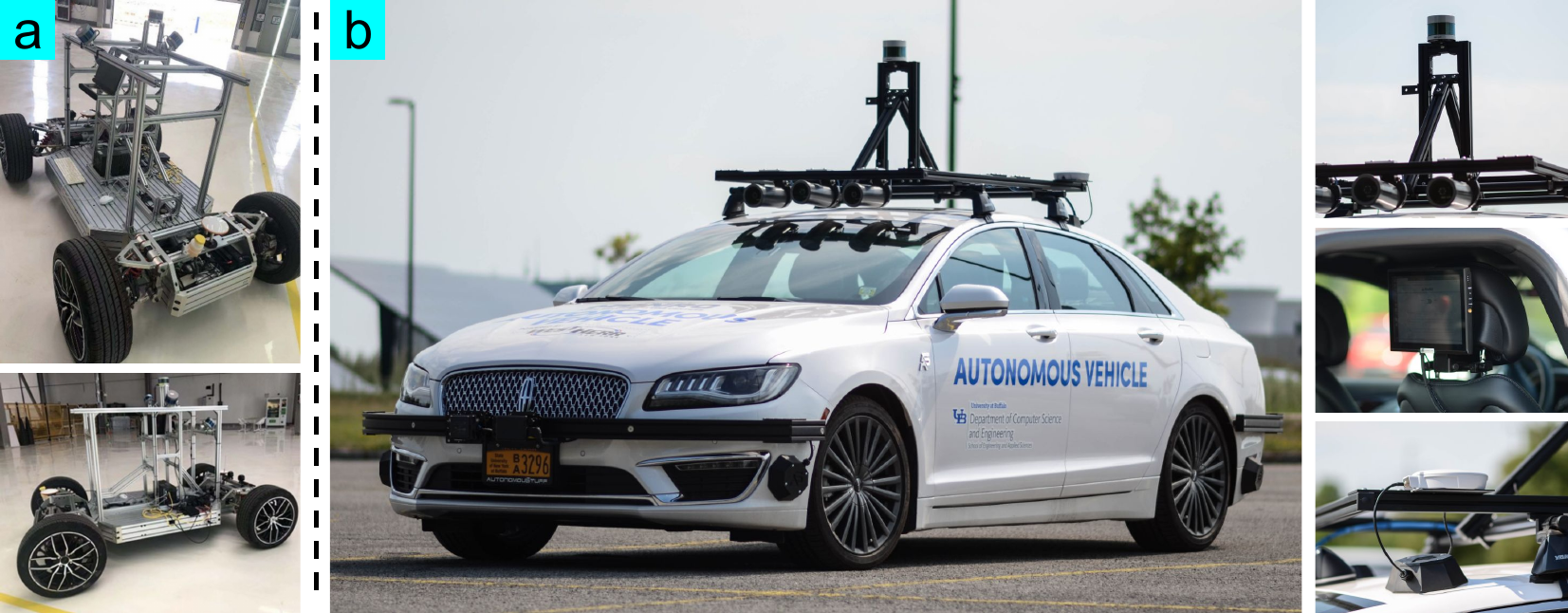}
\caption{AD vehicles for the system-driven evaluation platform: (a) A real-vehicle sized chassis with L4 AD sensors and closed-loop control; (b) A L4 AD vehicle built upon Lincoln MKZ. Both are equipped with L4 AD-grade sensors such as LiDARs, cameras, RADARs, GPS, and IMU. Detailed specifications are on our project website \textbf{\url{https://sites.google.com/view/cav-sec/pass}}~\cite{sok_website}.}
\label{fig:ubav}
\vspace{-0.1in}
\end{figure}

This is the testing vehicle and physical driving environment. As described in~\S\ref{sec:platform_design_goal_choices}, our platform is mainly built upon AD simulation (e.g., industry-grade ones such as SVL~\cite{svl}). To configure the simulation environment, we define a set of scenarios to describe the AD vehicle initial position, equipped sensors, testing road, and road objects (e.g., vehicles and pedestrians). The scenario descriptions are also provided as human-readable files for easy modification. Benefited from the general AD model structure, the evaluation can also be performed on a real AD vehicle (e.g., L4-capable vehicles as shown in Fig.~\ref{fig:ubav}) under the driving scenarios provided by the testing facility.

\subsubsection{Bridge} Bridge is the communication channel between the AD and plant models for sensor data reading and AD vehicle actuation. For better extensibility, it supports function hooking for modifying the communication data at runtime.

\subsubsection{Attack/Defense plugins} These two host the available attacks and defenses, which are built upon the attack/defense interfaces provided by the platform. To ease the deployment of future attacks/defenses, these interfaces are designed to support common attack vectors (\S\ref{sec:sok_attack_vectors}) and defense strategies (\S\ref{sec:sok_defense_methods}). Specifically, 3 general types of interfaces are defined based on the SoK:
\begin{itemize}
    \item \textit{Physical-world attack interface} enables dynamically loading 2D/3D objects to the simulation environment at arbitrary locations. It covers many physical-world attacks leveraging object texture, shape, and position.
    \item \textit{Sensor attack/defense interface} enables registering customized sensor data operations, which model the sensor attack or defense behaviors, in the bridge.
    \item \textit{Cyber-layer attack/defense interfaces} support adding/replacing components in the AD system and serve as versatile interfaces to enable cyber-layer attacks/defenses. For example, ML backdoor attacks can be implemented by replacing an existing ML model with a trojan one. Consistency-checking based defense can be implemented by adding component with consistency checking logic.
\end{itemize}

\subsubsection{Metric calculation} This is in charge of collecting measurements from all other modules in the platform and calculating the scenario-dependent evaluation metrics. Based on our SoK of attack goals (\S\ref{sec:sok_attack_goals}), we include system-level metrics such as safety (e.g., collision rate~\cite{jha2020ml, sato2021dirty}), traffic rule violation (e.g., lane departure rate~\cite{sato2021dirty, shen2020drift}), trip delay, etc. For comprehensiveness we also include component-level metrics (e.g., frame-wise attack success rate~\cite{zhao2019seeing, eykholt2018physical, chen2018shapeshifter}).

\textbf{Implementation.} We have implemented several variations for each module in the platform. The AD model supports modular AD pipelines for STOP sign attack evaluation (Appendix~\ref{appendix:example_platform_ad_plant_model_setups}) and full-stack AD systems (Apollo~\cite{github_apollo} and Autoware~\cite{github_autoware}). Our plant model includes an industry-grade AD simulator~\cite{svl} and real L4-capable AD vehicles (Fig.~\ref{fig:ubav}). For bridge, we reuse the ones (Python APIs, CyberRT, and ROS) provided by the AD simulator. Specifically, we modified the AD simulator, bridge, and modular AD pipeline to enable the 3 general attack/defense interfaces mentioned above. For metrics, we implemented collision and STOP sign-related violation checkings. Note that we \textit{do not intend to cover all existing attacks/defenses, and also believe we should not}; our goal (\S\ref{sec:platform_design_goal_choices}) is to initiate a community-level effort to \textit{collaboratively} build a common system-level evaluation infrastructure, which is the only way to make such community-level infrastructure support sustainable and up-to-date. We will fully open-source the platform (will be available on our project website~\cite{sok_website}) and welcome community contributions of new attack/defense interfaces and implementations.

\vspace{-0.1cm}
\nsubsection{Case Study: System-Level Evaluation on Stop Sign Attacks} \label{sec:platform_eval}

\textbf{Evaluation methodology.}
In this section, we use \textit{PASS} to perform system-level evaluations for the existing STOP sign attacks to showcase the platform capabilities and benefits. We choose STOP sign attacks because they are the most studied AD AI attack type in the literature (\S\ref{gap:attack_goals}) and \textit{none} of them have conducted system-level evaluations (\S\ref{sec:sok_eval_method}).

\textit{Evaluated attacks.} We evaluate the most safety-critical STOP sign AI attack goal: STOP sign \textit{hiding}, which generally leverage malicious object textures to make a STOP sign undetected (\S\ref{sec:sok_attack_vectors}). We reproduced 3 representative designs: \textit{ShapeShifter (SS)}~\cite{chen2018shapeshifter}, \textit{Robust Physical Perturbations (RP2)}~\cite{eykholt2018physical}, and \textit{Seeing Isn't Believing (SIB)}~\cite{zhao2019seeing}. Fig.~\ref{fig:stop_signs} (a) and (b) show the reproduced adversarial STOP signs. Details are in Appendix~\ref{appendix:attack_reproduction}.

\begin{figure}[tbp]
\centering
\includegraphics[width=\linewidth]{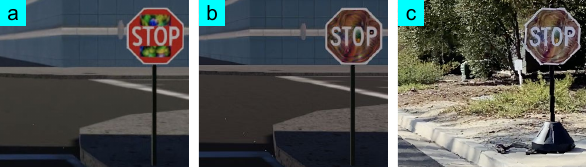}
\vspace{-0.2in}
\caption{Reproduced semantic AD AI attacks on STOP sign hiding: (a) RP2~\cite{eykholt2018physical} in simulator, (b) SS~\cite{chen2018shapeshifter} in simulator, and (c) SS~\cite{chen2018shapeshifter} in physical world.}
\label{fig:stop_signs}
\vspace{-0.2in}
\end{figure}

\textit{AD model configurations.} 
We configure the modular AD pipeline (\S\ref{sec:platform_design}) in the platform as a general AD system that includes camera object detection, object tracking, fusion (optional), planning, and control. Specifically, we configure 3 variations of the AD pipeline based on the availability of map and fusion (detailed configurations are in Appendix~\ref{appendix:example_platform_ad_plant_model_setups}).

\textit{Driving scenarios.} We select a straight urban road with a STOP sign at an intersection shown in Fig.~\ref{fig:stop_signs} (a) and (b). We evaluate under 5 common local-road driving speeds (10--30 mph with 5-mph step), 3 lighting conditions (sunrise, noon, sunset), and 3 weather conditions (sunny, cloudy, rainy). During evaluation, the malicious STOP sign images are deployed in the simulator via physical-world attack interface (\S\ref{sec:platform_design}).

\textit{Evaluation metrics.} To evaluate system-level attack effectiveness, we select traffic rule related metrics for STOP sign scenarios: \textit{stopping rate} refers to the percentage of successfully stopping (but may already violate the stop line), \textit{violation rate} is the percentage of violating the stop line, \textit{violation distance} is the distance over the stop line if violated. For comparison, we also calculate component-level metrics used in prior works~\cite{eykholt2018physical, zhao2019seeing}: \textit{frame-wise attack success rate}, $f_{\mathrm{succ}}$, is the percentage of frames that are misdetected in different STOP sign distance ranges, \textit{the best attack success rate in $n$ consecutive frames}, $f_{\mathrm{succ}}^{\mathrm{max}(n)}$, is proposed by Zhao et al.~\cite{zhao2019seeing} as a better metric to measure whether enough success has been accumulated. We use $n=50$ (instead of 100~\cite{zhao2019seeing}) since most of our simulation scenarios complete within 100 frames due to the low driving speeds (i.e., short braking distances).

\textbf{Result highlights.}
Our evaluation results show that at the component level, all 3 attacks have very high effectiveness:  SS achieves $f_{\mathrm{succ}} > 90\%$ in 10m and $f_{\mathrm{succ}}^{\mathrm{max}(50)}$ close to \textit{100\%} in all scenarios; RP2 and SIB both achieve $f_{\mathrm{succ}}^{\mathrm{max}(50)}=100\%$ when the speed is 10 mph. Both the aggregated attack results and those at specific distance/angle ranges are all very similar to the reported component-level attack effectiveness in the original papers (Appendix~\ref{appendix:attack_reproduction}). However, similar to prior security/robustness studies for other AD AI components~\cite{seshia2020semantic, jia2020fooling}, we find that such high attack success is \textit{far from achieving the system-level attack goals}: RP2 and SIB are not able to cause STOP sign violations at all across all driving scenario combinations; SS can only succeed when the speed is very low (10mph) while failing for all other speeds (15-30mph), which are more common speed ranges for STOP sign roads. More details are in Table~\ref{tab:allresults} in Appendix.

Such low system-level attack effectiveness is mainly due to the tracking component in the modular AD pipeline, which were set up using representative parameters recommended by Zhu et al.~\cite{zhu2018online}. Although the attacks are quite effective at close distances, the STOP sign track created before is still alive, so that the AD system is always aware of the STOP sign and keeps committing the stop decisions. For SS with low speed (10 mph), the attack can hide the detection in more consecutive frames, and thus the STOP sign track can be eventually deleted. This thus again validates the non-triviality of the AI-to-system semantic gap, which further demonstrates the necessity of system-level evaluations for semantic AD AI security (\S\ref{sec:gaps_system_level_eval}). Meanwhile, here we are able to experimentally quantify whether and how much the system-level attack effectiveness for a given AD AI attack depends on the driving scenario (i.e., driving speed), which is only made possible with our simulation-centric design that can more flexibly support different driving scenarios, AD designs, and system-level metrics.

\vspace{-0.1cm}
\nsubsection{Simulation Fidelity Evaluation} \label{sec:platform_eval_fidelity}

Although the simulation fidelity drawback is tolerable since (1) the simulation fidelity is evolving and can be improved by our real vehicle setup and (2) practical attacks/defenses should be capable of domain adaptations (\S\ref{sec:platform_design_goal_choices}), we still hope that today's simulation fidelity is readily usable for AD AI security evaluation purposes. In this section, we thus take the STOP sign attack as an example to concretely understand this.

\textbf{Experimental setup.} We use the SS attack, and calculate the similarity between the STOP sign detection results in the physical world and the simulation environment in our platform. The physical-world setup is shown in Fig.~\ref{fig:stop_signs} (c). Correspondingly, we use a simulation environment with similar road geometry. More details are in Appendix~\ref{appendix:simulation_fidelity_setup}.

\textbf{Results.} Among 20 physical-world driving traces, we find that an average correlation coefficient $r$ of 0.75 (lowest: 0.62, highest: 0.80), and all correlations are statistically significant ($p<0.05$). For Pearson's correlation, $r>0.5$ is considered strongly correlated~\cite{cohen2013statistical}. This shows that at least for such representative AD AI attack type, today's simulation fidelity is sufficient at least for our AD AI security evaluation purpose.

\vspace{-0.1cm}
\nsubsection{Educational Usage of \textit{PASS}} \label{sec:platform_ctf_leaderboard}

Beyond research usage, \textit{PASS} can also be used for educational purposes. We recently organize a Capture The Flag (CTF) event focusing on AD AI security (AutoDriving CTF~\cite{autodriving_ctf_website}, co-located with DEF CON 29). Specifically, we leverage an earlier version of \textit{PASS} to create simulation-based CTF challenges on GPS/LiDAR spoofing, lane detection attacks, etc. In total, over 100 teams across the globe participated in the CTF. Since \textit{PASS} was provided as a resource for the teams, after the event, we asked the 5 winning teams for feedback on the platform (IRB exempted). Particularly, all 5 teams consider the platform as helpful for solving the challenges. However, due to relatively high resource requirement (e.g., GPUs), 2 teams suggested providing the platform as an online service. This is also the reason why providing remote evaluation services is part of our design goals in~\S\ref{sec:platform_design_goal_choices}.
\nsection{Conclusion} \label{sec:conclusion}

In this paper, we perform the first systematization of knowledge of the growing semantic AD AI security research space. We collect and analyze 53 such papers, and systematically taxonomize them based on research aspects critical for the security field. We summarize 6 most substantial scientific gaps based on both quantitative vertically and horizontal comparisons, and provide insights and potential future directions not only at design level, but also at research goal, methodology, and community levels. To address the most critical scientific methodology-level gap, we take the initiative to develop an open-source, uniform, and extensible system-driven evaluation platform \textit{PASS} for the community. We hope that the systematization of knowledge, identified scientific gaps, insights, future direction, and our community-level evaluation infrastructure building efforts can foster more extensive, realistic, and democratized future research into this critical research space.

\bibliographystyle{ieeetr}
{
\footnotesize
\bibliography{survey, others}
}

\appendix
\nsubsection{Targeted AI Components} \label{appendix:sok_target_ai_components}

\begin{figure}[!htbp]
\centering
\vspace{-0.5cm}
\includegraphics[width=.9\linewidth]{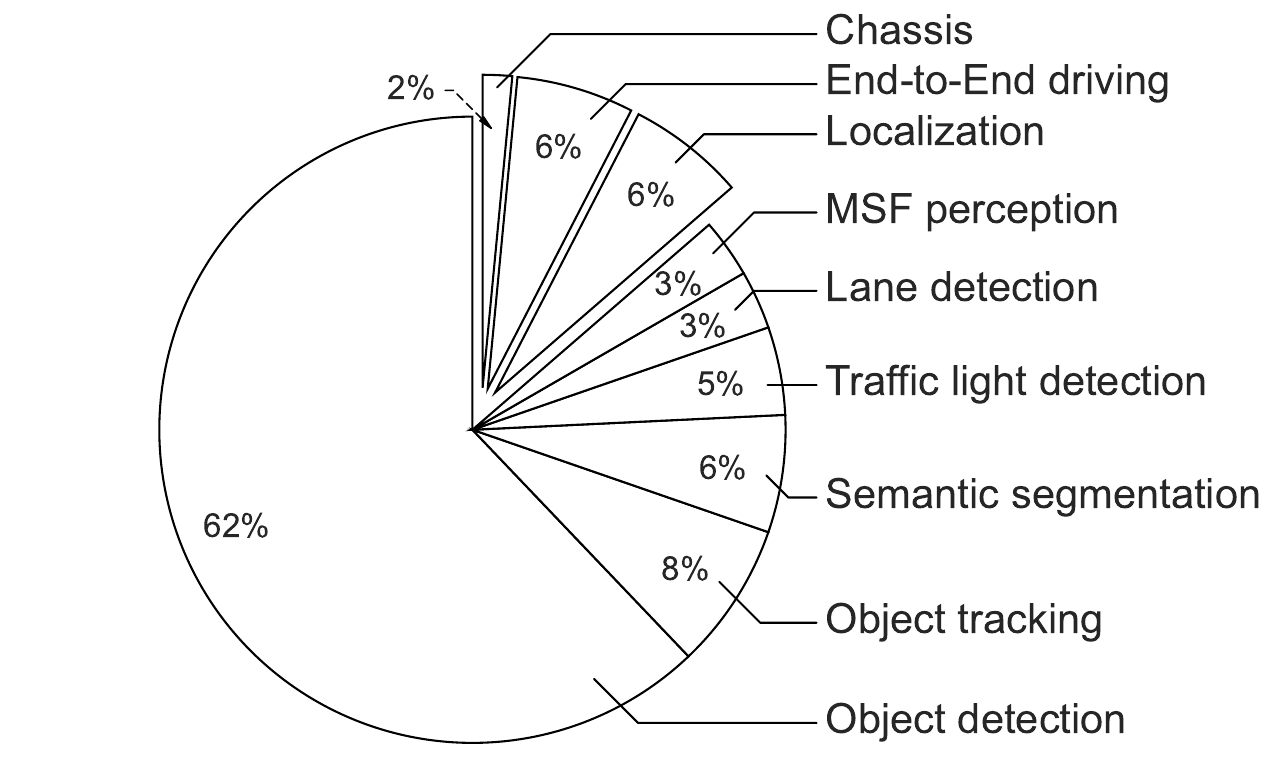}
\caption{Distribution of (attack/defense) targeted AI components in semantic AD AI security papers.}
\label{fig:stats_targeted_ai_components}
\end{figure}

\subsubsection{Perception} \label{sec:sok_perception} \ 

\textbf{Road object detection} identifies obstacles (e.g., bounding box), lane lines, and traffic lights from sensor data. The ones targeted by existing semantic AD AI security works includes:

\textit{Object detection} and \textit{segmentation} are commonly used to detect vehicles, pedestrians, cyclists, traffic objects (e.g., traffic cones), etc., in AD systems. Existing attacks on these aim to cause adversarial effects including \textit{hiding} (i.e., making objects disappear), \textit{creation} (i.e., detecting non-existing objects), and \textit{alteration} (i.e., changing the types of objects).

\textit{Lane detection} can detect lane line shapes and relative position in lane and is widely used in L2 AD systems for Automated Lane Centering. Prior works~\cite{sato2021dirty, jing2021too} show lane detection models are vulnerable to physical perturbations on road, which can lead AD vehicle to drive out of lane line.

\textit{Traffic light detection} identifies the traffic light and detects the current signal color. Prior works~\cite{tang2021fooling, lovisotto2021slap, wang2021i} leverages its dependency on localization or light projection to induce wrong detection, no detection, or color change.

\textit{MSF perception} is commonly used in L4 AD systems~\cite{github_apollo} to fuse detection results from different perception sources (e.g., LiDAR and camera) for higher accuracy and robustness~\cite{frossard2018end, liang2018deep, chen2017multi}. Prior works found that MSF perception is vulnerable to adversarial 3D objects~\cite{cao2021invisible, tu2021exploring}. Such attacks can hide the adversarial objects or cause other obstacles being misdetected. 

\textbf{Object tracking} builds the trajectories of one or more detected objects in a sensor frame sequence to tolerant the occasional false positions and false negatives. However, prior works~\cite{ding2021towards, chen2021unified, jia2020fooling} demonstrate successful adversarial attacks that can move a roadside vehicle into the driving lane or move a front vehicle out of the road~\cite{jia2020fooling, jha2020ml}.

\subsubsection{Localization} \label{sec:sok_localization} \

\textbf{LiDAR/Camera localizations} are commonly used in AD systems for localizing the vehicle. They operate by finding the best matched location of the live LiDAR point cloud or camera image in a map~\cite{ndt, mur2017orb}. Prior works discover that such localization algorithms may lead the sensitive location data~\cite{luo2020stealthy} or can be disrupted by IR lights~\cite{wang2021i}.

\textbf{MSF localization} is predominantly used in L4 AD systems to achieve robust and accurate localization by fusing location sources such as GPS, LiDAR, and IMU~\cite{wan2018robust}. Prior work~\cite{shen2020drift} finds that Kalman Filter based MSF design is vulnerable to strategic attacks leveraging a single attack source, such as GPS spoofing, to inject large deviations in the localization results.

\subsubsection{Chassis} \label{sec:sok_chassis} The Chassis component acts as the information hub of the vehicle, which periodically broadcast privacy-sensitive diagnosis information, including Vehicle Identification Number (VIN), to other components. Prior work~\cite{hong2020avguardian} demonstrates using a compromised Robot Operating System (ROS) node to intercept the Chassis message to steal the VIN.

\subsubsection{End-to-End Driving} \label{sec:sok_e2e} End-to-End driving~\cite{tampuu2020survey} is a distinctive AD system design paradigm from the more common modular designs used in industry-grade AD systems~\cite{github_apollo, github_autoware}. Due to DNN-based design, end-to-end driving is vulnerable to adversarial attacks, which can cause driving model to predict incorrect control commands~\cite{kong2020physgan, hamdi2020sada}. Prior work~\cite{liu2018trojaning} show that  trojan attack enables triggering specific driving behaviors (e.g., turn right) using road signs with specific patterns.

\vspace{-0.2cm}
\nsubsection{STOP Sign Attack Reproduction} \label{appendix:attack_reproduction}

\textbf{ShapeShifter (SS)~\cite{chen2018shapeshifter} reproduction and results.} We directly use the official open-source code~\cite{ssgithub} to reproduce SS on Faster R-CNN~\cite{ren2015faster}. With that, we can ensure that the attack generated by the source code should have the same effect as the original work. We validated in simulation using the same distance/angle settings as the paper. Results show that our reproduction can achieve 80\% (12/15) attack success rate while the original paper reports 73\% (11/15). In addition, the successful attack distances/angles of the reproduced attack align well with the paper with the only exception that our reproduction is successful at 25 feet / 0$^{\circ}$ but the paper reports failure. Such results also indicate a sufficient simulation fidelity, which is consistent with our fidelity evaluation in \S\ref{sec:platform_eval_fidelity}.

\textbf{Robust Physical Perturbation (RP2)~\cite{eykholt2018physical} reproduction and results.}  Same as the paper, we select YOLOv2~\cite{redmon2017yolov2} as the targeted object detection model. Specifically, we implement the loss function including adversarial loss with Expectation over Transformation~\cite{athalye2018synthesizing}, total variation loss, $l_p$ loss, and non-printability score loss~\cite{sharif2016accessorize} as used in RP2 design~\cite{eykholt2018physical}. We compare the attack effectiveness in simulation to the reported results in the paper in the same distance settings (0--30 feet in outdoor environment). Our results show that we can achieve an attack success rate of 66.4\% in all camera frames, which is similar to the 63.5\% reported in the paper.

\textbf{Seeing Isn't Believing (SIB)~\cite{zhao2019seeing} reproduction/modeling.} Since SIB is not open-sourced, we tried our best to reproduce it but were still unable to achieve the same attack effectiveness shown in their paper~\cite{zhao2019seeing}. Nevertheless, in our simulation platform evaluation (\S\ref{sec:platform_eval}), we apply SIB using a modeling-based approach, i.e., by sampling STOP sign detection failures using \textit{exactly the same} frame-wise attack success rate at different STOP sign distances/angles reported in the paper. Specifically, the attack success rates in the paper are from 32--94\% for distances from 5--25 m on YOLOv3~\cite{zhao2019seeing}.

\vspace{-0.2cm}
\nsubsection{Example AD and Plant Model Setups in PASS} \label{appendix:example_platform_ad_plant_model_setups}

\textbf{AD model setup.} We illustrate the detailed AD model setup for the STOP sign attack evaluation (\S\ref{sec:platform_eval}) as an example to demonstrate the usage of \textit{PASS}. Specifically, we adopt the typical modular AD pipeline design including object detection, object tracking, fusion, planning, and control.

\textit{Object detection.} We use the same models and detection thresholds as originally used in the STOP sign attacks: SS on Faster R-CNN (threshold: 0.3)~\cite{chen2018shapeshifter}, RP2 on YOLOv2 (threshold: 0.4)~\cite{eykholt2018physical}, and SIB on YOLOv3 (threshold: 0.4)~\cite{zhao2019seeing}. 

\textit{Object tracking.} We adopt a general Kalman Filter based multi-object tracker~\cite{luo2020multiple} which tracks the STOP sign locations and sizes. Next, we convert the detected STOP sign location from 2D image coordinates to real-world coordinates. Here, we adopt 2 variants: (1) HD Map based (denoted \textit{Map}). Similar to the design in Tesla Autopilot~\cite{tesla_stop_traffic_sign_control}, we use the GPS position to query the HD Map for the distance to the front intersection and use it as an approximation for the STOP sign distance. (2) Pinhole camera based (denoted \textit{Pinhole}). We apply the pinhole camera model to estimate the STOP sign distance based on the heuristics on standard STOP sign sizes~\cite{road_sign_sizes}, which is a similar design to Apollo~\cite{apollo_pinhole_camera}. For track management, a new STOP sign track is created when the STOP sign is detected for 0.2 $\times$ FPS = 4 (frames per second) frames, and an existing track will be deleted after misdetection for 2 $\times$ FPS = 40 consecutive frames (parameters recommended by Zhu et al.~\cite{zhu2018online}). 

\textit{Fusion.} The fusion component is optional and thus can be disabled in the pipeline. If enabled, it can fuse object detection results from multiple data sources. In addition, if the HD Map contains detailed static road objects (e.g., traffic signs) it can serve as a data source to fuse with object detection model outputs. Since the output format of all fusion sources are consistent (i.e., objects), we extend the object tracker to accept detections from multiple sources to facilitate the fusion.

\textit{Planning.} We adopt a lane following planner where the future trajectory locates at the center of current driving lane~\cite{github_apollo}. The planner by default maintains a desired speed, but reduces it if \textit{a STOP sign presents} or \textit{there exists a front obstacle with a close distance or slowing down}. Taking the STOP sign as an example, we first calculate a \textit{braking distance} based on the current driving speed and the safe vehicle deceleration ($-3.4 m/s^2$~\cite{deceleration}). If a STOP sign is currently tracked and its distance to the vehicle reaches the braking distance, the planner starts to decelerate. If the STOP sign is no longer tracked and current speed is smaller than the desired driving speed (i.e., the speed that the vehicle intends to maintain if no STOP sign), the planner accelerates the vehicle. 

\textit{Control.} The control module will execute the planning decisions, and we use the classic Stanley controller~\cite{hoffmann2007autonomous} for lateral control (i.e., steering) and a Proportional-Integral-Derivative (PID) controller~\cite{ang2005pid} for longitudinal control (i.e., throttling and braking). Consistent with existing L2 AD systems such as OpenPilot~\cite{news_openpilot}, 
We set the frequency of detection, tracking, fusion, and planning as 20 Hz, and control as 100 Hz. 

In our evaluation (\S\ref{sec:platform_eval}), We adopt 3 AD pipeline variants based on the availability of HD Map and fusion: \textit{Map}, \textit{Pinhole}, and \textit{Fusion}. Specifically, \textit{Map} and \textit{Pinhole} vary in the STOP distance estimation, and \textit{Fusion} refers to fusing STOP sign detection with the one from the HD Map.

\textbf{Plant model setups.} \textit{PASS} supports two options for the plant model. For simulation-based evaluation, we adopt the SVL~\cite{svl}, which is an industry-grade AD simulator. For real-world experiments, we can use the two L4-capable AD vehicles in our project team as shown in Fig.~\ref{fig:ubav}, where they are equipped with AD-grade sensors including multiple short- and long-range cameras, LiDARs (16-line, 32-line, and 64-line), mmWave RADAR, duel-antennas GPS with RTK, high-precision IMU, etc. As described in \S\ref{sec:platform_design_goal_choices}, the platform is simulation-centric; it is mainly for simulation-based evaluation and only use the real vehicles for simulation fidelity improvements and physical world evaluation in exceptional cases.

\nsubsection{Setup for Simulation Fidelity Evaluation} \label{appendix:simulation_fidelity_setup}

In the physical-world experiments, we print the adversarial STOP sign and overlay it on a portable STOP sign as shown in Fig.~\ref{fig:stop_signs} (c). We have obtained the permit from our institute's transportation department to reserve a dead-end single-lane road for controlled experiments. Correspondingly, we use a simulation environment with similar road geometry. In both settings, the vehicle drives at 10 mph from 300 feet until passes the STOP sign. In the physical-world setting, we use an iPhone 11 mounted on the vehicle windshield to record the driving traces. We collect a total of 20 driving traces with different STOP sign placements in the physical world. Since simulation results are generally quite stable, we only collect one trace in the simulation. We then run object detection on each camera frame to obtain the STOP sign detection confidences in the driving traces. To ensure the traces are synchronized and comparable, we trim and interpolate confidences in the physical-world traces based to a 20 Hz frequency same as in the simulation. We then calculate the Pearson's correlation between the simulation trace and \textit{each} physical-world trace.

\begin{table}[]
\centering
\setlength{\tabcolsep}{1.2pt}
\begin{tabular}{|c?c?c?c?c|ccc?c|ccc|}
\hline
\multicolumn{1}{|c?}{} & \multicolumn{1}{c?}{} & \multicolumn{1}{c?}{} & \multicolumn{1}{c?}{} &  & \multicolumn{4}{c|}{Component-level metrics} & \multicolumn{3}{c|}{System-level metrics} \\ \cline{6-12} 
\multicolumn{1}{|c?}{} &  & \multicolumn{1}{c?}{Spd.} & \multicolumn{1}{c?}{} &  & \multicolumn{3}{c?}{$f_{\mathrm{succ}}$} & \multirow{2}{*}{\begin{tabular}[c]{@{}c@{}}$f_{\mathrm{succ}}^{\mathrm{max}(50)}$\end{tabular}} & Stop & Vio. & Vio. \\ \cline{6-8}
\multicolumn{1}{|c?}{\multirow{-3}{*}{\begin{tabular}[l]{@{}c@{}}\vthead{Attack}\end{tabular}}}& \multicolumn{1}{c?}{\multirow{-3}{*}{\begin{tabular}[l]{@{}c@{}}\vthead{AD Pip}\end{tabular}}} & \multicolumn{1}{c?}{(mph)} & \multicolumn{1}{c?}{L} & W & \textless{}10m & \textless{}20m  & \textless{}30m &  & rate & rate & dist. \\
\hline
\multirow{19}{*}{\begin{tabular}[c]{@{}c@{}}\vthead{SS~\cite{chen2018shapeshifter}}\end{tabular}} & \multirow{9}{*}{\begin{tabular}[c]{@{}c@{}}\vthead{Map}\end{tabular}} & 10 &  &  & 91.3\% &  78.8\% & 67.4\%  & 100.0\%  & \textbf{0.0\%} & \textbf{100.0\%}  & \textbf{$\infty$} \\ 
 &  & 15  &  &  & 99.8\%   & 82.2\%  & 60.4\%   & 99.8\%  & 100.0\% & 0.0\%  & 0 m \\ 
 &  & 20  & N & S & 100.0\%  & 87.8\% & 68.6\%   & 100.0\%  & 100.0\% & 0.0\%  & 0 m \\ 
 &  & 25  &  &  & 100.0\% &  88.0\%  & 69.2\%  & 100.0\%  & 100.0\% & 0.0\%  & 0 m \\ 
 &  & 30  &  &  & 100.0\% &  88.7\%  & 71.2\%  & 98.0\%  & 100.0\% & 0.0\%  & 0 m \\ \cline{3-12}
  &  & \multirow{4}{*}{\begin{tabular}[c]{@{}c@{}}25 \end{tabular}} & \multirow{2}{*}{\begin{tabular}[c]{@{}c@{}}N\end{tabular}} & C & 100.0\%  & 88.8\%  & 69.8\%   & 100.0\%  & 100.0\% & 0.0\%  & 0 m \\ 
 &  &  &  & R & 99.3\% & 86.7\%  & 68.3\% & 99.0\%  & 100.0\% & 0.0\%  & 0 m \\ \cline{4-12}
 &  &  & SR & \multirow{2}{*}{\begin{tabular}[c]{@{}c@{}}S\end{tabular}} & 100.0\% &  83.9\%  & 66.0\% &   100.0\%  & 100.0\% & 0.0\%  & 0 m \\ 
 &  &  & SUS &  & 100.0\% &  88.0\%  & 69.2\%  & 100.0\%  & 100.0\% & 0.0\%  & 0 m \\ \cline{2-12}
 & \multirow{5}{*}{\begin{tabular}[c]{@{}c@{}}\vthead{Pinhole}\end{tabular}} & 10  &  &  & 91.3\% &  78.8\%   &  67.5\% & 100.0\%  & \textbf{0.0\%} & \textbf{100.0\%}  & \textbf{$\infty$} \\ 
 &  & 15  &  &  & 100.0\%  & 82.5\%  & 60.6\%  & 100.0\%  & 100.0\% & 0.0\%  & 0 m \\ 
 &  & 20  & N & S & 100.0\%  & 87.8\%  & 68.6\%  & 100.0\%  & 100.0\% & 0.0\%  & 0 m \\ 
 &  & 25  &  &  & 100.0\%  & 88.5\%  & 69.5\%  & 100.0\%  & 100.0\% & 0.0\%  & 0 m \\ 
 &  & 30  &  &  & 100.0\%  & 85.4\%  & 68.6\%  & 97.1\%  & 100.0\% & 0.0\%  & 0 m \\  \cline{2-12}
 & \multirow{5}{*}{\begin{tabular}[c]{@{}c@{}}\vthead{Fusion}\end{tabular}} & 10  &  &  & 100.0\%  & 86.1\%  &  73.8\% & 100.0\%  & 100.0\% & 0.0\%  & 0 m \\ 
 &  & 15  &  &  & 100.0\%  & 82.3\%  & 60.4\%  & 100.0\%  & 100.0\% & 0.0\%  & 0 m \\ 
 &  & 20  & N & S & 100.0\%  & 87.8\%  & 68.6\% & 100.0\%  & 100.0\% & 0.0\%  & 0 m \\ 
 &  & 25  &  &  & 100.0\%  & 89.8\%  & 70.6\%  & 99.8\%  & 100.0\% & 0.0\%  & 0 m \\ 
 &  & 30  &  &  & 100.0\%  & 88.7\%  & 71.2\%  & 97.5\%  & 100.0\% & 0.0\%  & 0 m \\  \hline
 \multirow{19}{*}{\begin{tabular}[c]{@{}c@{}}\vthead{RP2~\cite{eykholt2018physical}}\end{tabular}}& \multirow{9}{*}{\begin{tabular}[c]{@{}c@{}}\vthead{Map}\end{tabular}} & 10  &  &  & 78.0\%  & 51.8\%  & 46.3\% & 100.0\%  & 100.0\% & 0.0\%  & 0 m \\ 
 &  & 15  &  &  & 73.4\%  & 46.9\%  & 46.0\%  & 84.5\%  & 100.0\% & 0.0\%  & 0 m \\ 
 &  & 20  & N & S & 76.1\%  & 54.8\%  & 52.2\%  & 88.0\%  & 100.0\% & 0.0\%  & 0 m \\ 
 &  & 25  &  &  & 61.6\%  & 42.1\%  & 43.6\%   & 54.5\%  & 100.0\% & 0.0\%  & 0 m \\ 
 &  & 30  &  &  & 70.0\%  & 49.8\%  & 45.8\%  & 59.6\%  & 100.0\% & 0.0\%  & 0 m \\ \cline{3-12}
 &  & \multirow{4}{*}{\begin{tabular}[c]{@{}c@{}}25 \end{tabular}} & \multirow{2}{*}{\begin{tabular}[c]{@{}c@{}}N\end{tabular}} & C & 56.7\%  & 38.6\%  & 36.8\% & 50.0\%  & 100.0\% & 0.0\%  & 0 m \\ 
 &  &  &  & R & 66.6\%  & 49.4\%  & 54.2\%   & 65.5\%  & 100.0\% & 0.0\%  & 0 m \\ \cline{4-12}
 &  &  & SR & \multirow{2}{*}{\begin{tabular}[c]{@{}c@{}}S\end{tabular}} & 59.5\%  & 44.5\%  & 54.5\%  & 71.6\%  & 100.0\% & 0.0\%  & 0 m \\ 
 &  &  & SUS &  & 63.8\%  & 48.3\%  & 54.6\%  & 69.0\%  & 100.0\% & 0.0\%  & 0 m \\ \cline{2-12}
 & \multirow{5}{*}{\begin{tabular}[c]{@{}c@{}}\vthead{Pinhole}\end{tabular}} & 10  &  &  & 78.0\% & 51.8\%   &  46.3\% & 100.0\%  & 100.0\% & 0.0\%  & 0 m \\ 
 &  & 15  &  &  & 73.6\%  & 47.0\%  & 46.4\%  & 85.1\%  & 100.0\% & 0.0\%  & 0 m \\ 
 &  & 20  & N & S & 77.7\%  & 56.8\%  & 53.9\%  & 91.4\%  & 100.0\% & 0.0\%  & 0 m \\ 
 &  & 25  &  &  & 60.9\% & 42.4\%  & 44.0\%   & 54.9\%  & 100.0\% & 0.0\%  & 0 m \\ 
 &  & 30  &  &  & 74.5\%  & 53.8\%  & 49.1\%  & 64.3\%  & 100.0\% & 0.0\%  & 0 m \\  \cline{2-12}
 & \multirow{5}{*}{\begin{tabular}[c]{@{}c@{}}\vthead{Fusion}\end{tabular}} & 10  &  &  & 78.0\% & 51.8\%    & 46.3\% & 100.0\%  & 100.0\% & 0.0\%  & 0 m \\ 
 &  & 15  &  &  & 73.4\%  & 46.9\% & 46.0\%  & 84.5\%  & 100.0\% & 0.0\%  & 0 m \\ 
 & & 20  & N & S & 78.0\%  & 56.1\%  & 53.3\%  & 90.2\%  & 100.0\% & 0.0\%  & 0 m \\ 
 &  & 25  &  &  & 66.3\%  & 46.2\%  & 47.0\% & 59.8\%  & 100.0\% & 0.0\%  & 0 m \\ 
 &  & 30  &  &  & 76.2\%  & 55.1\%  & 49.6\%  & 65.9\%  & 100.0\% & 0.0\%  & 0 m \\   \hline
\multirow{10}{*}{\begin{tabular}[c]{@{}c@{}}\vthead{SIB~\cite{zhao2019seeing}}\end{tabular}} & \multirow{5}{*}{\begin{tabular}[c]{@{}c@{}}\vthead{Map}\end{tabular}} & 10  &  &  & 74.7\%  & 57.4\%   & 46.7\% & 100.0\%  & 100.0\% & 0.0\%  & 0 m \\ 
 &  & 15  &  &  & 45.3\%  & 28.9\%  & 21.2\% & 47.1\%  & 100.0\% & 0.0\%  & 0 m \\ 
 & & 20  & N & S & 73.7\%  & 59.8\%  & 50.3\%  & 100.0\%  & 100.0\% & 0.0\%  & 0 m \\ 
 &  & 25  &  &  & 76.7\%  & 66.2\%  & 58.9\%  & 100.0\%  & 100.0\% & 0.0\%  & 0 m \\ 
 &  & 30  &  &  & 83.4\%  & 74.1\%  & 67.8\%  & 100.0\%  & 100.0\% & 0.0\%  & 0 m \\  \cline{2-12}
 & \multirow{5}{*}{\begin{tabular}[c]{@{}c@{}}\vthead{Fusion}\end{tabular}} & 10  &  &  & 74.7\%  & 57.4\%    & 46.7\% & 100.0\%  & 100.0\% & 0.0\%  & 0 m \\ 
 &  & 15  &  &  & 45.3\%  & 28.9\% & 21.2\% & 47.1\%  & 100.0\% & 0.0\%  & 0 m \\ 
 & & 20  & N & S & 73.7\%  & 59.8\% & 50.3\%& 100.0\%  & 100.0\% & 0.0\%  & 0 m \\ 
 &  & 25  &  &  & 76.7\%  & 66.2\% &  59.0\% & 100.0\%  & 100.0\% & 0.0\%  & 0 m \\ 
 &  & 30  &  &  & 80.6\%  & 70.4\%  & 63.8\% & 97.5\%  & 100.0\% & 0.0\%  & 0 m \\ \hline
\end{tabular}

\vspace{0.05in}
\footnotesize
SR, N, SUS: sunrise, noon, sunset (6 am, 12 pm, 18 pm), C: cloudy (SVL cloudiness = 0.5), R: rainy (SVL rain = 0.5); \\
AD Pip: AD pipeline variants, Spd.: speed, L: lighting, W: weather, \\ vio.: violation, dist.: distance;\\
$f_{\mathrm{succ}}^{\mathrm{max}(50)}$: best attack success rate in 50 consecutive frames;\\
$f_{\mathrm{succ}}$: frame-wise attack success rate.

\caption[overview]{Component- and system-level evaluation results of the 3 stop sign AI attacks on our evaluation platform. The results are averaged over 10 runs of each configuration. Detailed AD pipeline setup can be found in Appendix~\ref{appendix:example_platform_ad_plant_model_setups}. We do not apply the lighting and weather variations for SIB since the object detection outputs are directly modeled and thus will not be influenced by them (Appendix~\ref{appendix:attack_reproduction}).
}
\vspace{-0.4cm}
\label{tab:allresults}
\end{table}

\end{document}